\documentclass[12pt]{JHEP3}
\usepackage{epsfig}
\usepackage{amsmath}
\def\beq{\begin{eqnarray}}
\def\eeq{\end{eqnarray}}
\newcommand{\Ncal}{\ensuremath{\mathcal{N}}}

\title{
Toward a Proof of Montonen-Olive Duality via Multiple M2-branes
}
\author{
Koji Hashimoto$^{\dagger a}$, 
Ta-Sheng Tai$^{\dagger b}$
and 
Seiji Terashima$^{* c}$\\
${}^\dagger$ {\it Theoretical Physics Laboratory, 
Nishina Center, \\
\hspace{40mm} RIKEN, Saitama 351-0198, Japan }\\
${}^*$ {\it Yukawa Institute for Theoretical Physics, \\
\hspace{40mm} Kyoto University, Kyoto 606-8502, Japan}\\
$^a$ E-mail: \email{koji@riken.jp}\\
$^b$ E-mail: \email{tasheng@riken.jp}\\
$^c$ E-mail: \email{terasima@yukawa.kyoto-u.ac.jp}\\
}

\abstract
{
We derive 4-dimensional
${\cal N}=4$ $U(N)$ supersymmetric Yang-Mills theory
from a 3-dimensional Chern-Simons-matter
theory with product gauge group $(U(N))^{2n}$.
The latter describes M2-branes probing an orbifold
where a torus emerges in a scaling limit.
It is expected that
the $SL(2,{\bf Z})$ duality of the 4-dimensional Yang-Mills theory
will be shown in M-theory point of view
since it is trivially realized as modular transformations of the torus.
Indeed, starting from one single Chern-Simons-matter theory,
we find infinitely many equivalent
4-dimensional
theories differing up to $T$-transformation
of the $SL(2,{\bf Z})$ redefinition of
the gauge coupling $\tau=\frac{\theta}{2\pi} + \frac{4\pi i}{g^2}$
and a parity transformation in 4 dimensions.
Although $S$-transformation can not be shown in our work,
it is important that
a part of the $SL(2,{\bf Z})$ transformation is realized via
the M2-brane action.
Thus we think our work can be a step
toward a proof of Montonen-Olive duality via M2-branes.}

\preprint{
{\normalsize RIKEN-TH-136}\\
{\normalsize YITP-08-75}
}

\begin{document}

\section{Introduction : A path to Montonen-Olive duality}
\label{section1}

Among recent developments on effective
actions of multiple M2-branes, 
one of the surprising outputs is the non-abelian duality. In 3 dimensions,
it has been known for more than a decade that the action 
of a single M2-brane 
can be dualized classically to that of a D2-brane 
\cite{Bergshoeff:1996tu}: 
an 
abelian duality between scalar field theory and 3d electromagnetism. 
Based on Bugger-Lambert-Gustavsson 
(BLG) model \cite{Bagger:2007jr, Gustavsson:2007vu} of multiple M2-brane, Mukhi and 
Papageorgakis have obtained 
a quite intriguing non-abelian duality 
\cite{Mukhi:2008ux}, a relation between field theories on multiple
M2-branes and D2-branes. 
In this paper, we generalize their result and find a novel 
mechanism which will be a step toward a proof of  
renowned Montonen-Olive (MO) duality conjectured \cite{Montonen:1977sn}
for 4d ${\cal N}=4$ $U(N)$ supersymmetric Yang-Mills (SYM) theory. 

The method of \cite{Mukhi:2008ux} 
is as follows. 
First, one of the eight transverse fields is given a vacuum expectation
value (vev) $v$,  
which turns out to provide mass terms for a non-dynamical part of 
the gauge fields in Chern-Simons (CS) terms. Integrating massive modes 
out, 
one gets rightly a YM kinetic term of D2-branes.

As pioneered by Aharony, Bergman, Jafferis and Maldacena (ABJM) 
\cite{Aharony:2008ug}, 
following the renormalization group (RG), certain ${\Ncal=3}$ 3d CSYM
theories 
flow to ${\Ncal=6}$ superconformal field theories (SCFTs) 
at IR fixed point, 
which precisely describe M2-branes probing 
${\bf C}^4/{\bf Z}_k$ $(k>2)$ where $k$ is the CS level.
Equipped with these, the need for a vev then gets clarified
geometrically. Under the scaling limit \cite{Distler:2008mk}:  
\begin{eqnarray}
v\rightarrow \infty,\quad k\rightarrow\infty,\quad v/k: \mbox{  fixed} , 
\label{limvk}
\end{eqnarray}
one yields exactly a circle compactification,
 {\it i.e.}~the M-theory circle is created in the above limit and 
D2-branes appear thereof. 
It is highly non-trivial 
that this mechanism requires CS terms which in turn give 
an orbifold moduli space.  

Let us extend  
the step further to 4-dimensional theories. 
Our result shows a perfect consistency with M-theory considerations. 
Generally, MO duality 
changes the gauge group, but in the case of $U(N)$ it remains the 
same.\footnote{For an introduction, 
see review articles \cite{Harvey:1996ur} and 
references therein.}

Let us summarize how the ${\Ncal=4}$ SYM, the M2-branes and
MO duality are related to each
other.  
As is well known, the axio-dilaton $\tau$ of 
Type IIB supergravity coincides with the 
gauge coupling of the ${\Ncal =4}$ SYM, realized on 
$N$ coincident D3-branes at low energy. MO duality is then 
identified with the $S$-duality
of Type IIB string theory.
In terms of M-theory, 
$\tau$ is interpreted as the complex structure of a two-torus formed by
$(x^9, x^{11})$ such that 
the $S$-duality gets readily identified with the 
$SL(2,{\bf Z})$ modular transformation. Also, via duality chains, 
M2-branes 
transverse to the above two-torus with shrinking size and fixed 
$\tau$ guarantee that the above D3-branes they are dual to are non-compact.

To get a torus, it is 
insufficient to just dwell in the present ABJM model. 
We make use of 
a generalized version studied by 
\cite{Benna:2008zy,Terashima:2008ba,Imamura:2008dt}.
It is this CS-matter theory with 
product gauge group $(U(N))^{2n}$
that comes to our rescue. 
The standard orbifolding action of Douglas and Moore 
\cite{Douglas:1996sw}
has been applied to the ABJM model \cite{Terashima:2008ba}
to obtain the generalized action.\footnote{
Orbifolding the ABJM model was first considered in
\cite{Fuji:2008yj}.}
As shown in \cite{Imamura:2008dt}, 
it is also possible to prepare a IIB brane setup which flows to 
the same theory at IR fixed point. Viewed from M-theory, 
this describes $N$ M2-branes probing an abelian orbifold ${\bf C}^4/\Gamma$ where 
$\Gamma={\bf Z}_n \times {\bf Z}_{kn}$. 

We then turn on vevs $(v,\tilde{v})$ of
two scalars and make a torus using instead the scaling 
limit:
\begin{align}
\begin{aligned}
&& v,\; \tilde{v},\; n \to \infty, \quad\quad
v\tilde{v}/{n} \to 0, \quad\quad
  v/\tilde{v} : {\rm fixed},
\quad k : {\rm fixed}, 
\label{limitnn}
\end{aligned}
\end{align}
which carries out the shrinking size with fixed $\tau$. 
Our field theory result shows that 
a 4d SCFT (SYM theory) emerges from a 3d SCFT  
(generalized ABJM model) as desired. Methods for uplifting  dimensions
are basically Taylor's T-duality 
\cite{Taylor:1996ik} and deconstruction of extra dimensions 
\cite{ArkaniHamed:2001ca}. See Sec.~\ref{section3} 
for details. 

With the SYM obtained, we go further to analyze MO duality. 
It is found that there are infinitely many equivalent SYMs derived from 
the same generalized ABJM action. But eventually they differ merely 
up to $SL(2,{\bf Z})$ redefinition of $\tau$.
This provides a proof of duality under some $SL(2,{\bf Z})$
transformations.  
Relabeling CS gauge fields gives rise to many SYMs.
In particular, one relabeling is to exchange $v$
and $\tilde{v}$, {\it i.e.} this resembles the M-theory 9-11 flip,
as two vevs create the torus. 

However, $\tau$ in fact 
depends only on the ratio $v/\tilde{v}$, which forms 
a one-parameter
family. This constraint rather implies that  
the $S$-transformation $(\tau \to -\tau^{-1})$ found 
between gauge couplings can be thought of as a 
parity transformation in 4 dimensions.
Thus generic (truely strong-weak dual) $S$-transformations 
are not shown in our work.
It is, however, important that
a part of the $SL(2,{\bf Z})$ transformation is realized in the field
theoretical framework via
the M2-brane action.
Thus we think our work can be a step
toward a proof of the MO duality via M2-branes.

The organization of this paper is as follows. In the next section we
briefly review the non-abelian duality of \cite{Mukhi:2008ux} 
and apply it to the ABJM model.
Then, Sec.~\ref{section3} is devoted to deriving 4d SYM
(\ref{ym4recom})  
from the generalized ABJM action (\ref{abjm2}). 
In Sec.~\ref{section4}, we obtain infinitely many SYMs 
from one parent model and find their gauge couplings are related by 
$SL(2,{\bf Z})$ transformation. This manifests MO duality henceforth. 
Certainly, our gauge coupling is in perfect accordance with   
the M-theory picture. 
Finally, we conclude in Sec.~\ref{section5}.

\section{Review : Scaling limit of orbifold and $S^1$ compactification}
\label{section2}
\setcounter{footnote}{1}

The ABJM model \cite{Aharony:2008ug} is a 3d ${\cal N}=6$ 
$U(N)\times U(N)$ CS-matter 
theory. It is conjectured to describe
$N$ M2-branes probing ${\bf C}^4/{\bf Z}_k$. 
In this section, after reviewing the ABJM model, we discuss the relation
between the scaling limit of an orbifold and the circle
compactification. This interesting mechanism giving a 
3d YM theory is {\it a la} Mukhi {\it et al.}
\cite{Mukhi:2008ux,Distler:2008mk}. 

An ultraviolet Type IIB brane configuration 
realizing $\Ncal=3$ $U(N)\times U(N)$ quiver YMCS theory in 3 dimensions is 
first given by \cite{Aharony:2008ug}. 
At low energy, it flows to a strongly-coupled $\Ncal=6$ SCFT. 
The bosonic part of the ABJM action is  
\begin{eqnarray}
S = \int \! d^3x 
\left[
\frac{k}{4\pi}\epsilon^{\mu\nu\lambda}
{\rm tr}
\left(
A^{(1)}_\mu \partial_\nu A^{(1)}_\lambda + \frac{2i}{3}A^{(1)}_\mu
A^{(1)}_\nu A^{(1)}_\lambda 
-
A^{(2)}_\mu \partial_\nu A^{(2)}_\lambda - \frac{2i}{3}A^{(2)}_\mu
A^{(2)}_\nu A^{(2)}_\lambda 
\right)
\right.
\nonumber \\
\left.
-{\rm tr} \left(
(D_\mu Z_A)^\dagger D^\mu Z^A
\right)
-{\rm tr} \left(
(D_\mu W^A)^\dagger D^\mu W_A
\right) - V(Z,W)
\right], \hspace{20mm}
\label{abjm}
\end{eqnarray}
where $A=1,2$, and kinetic terms of adjoint fields decouple due to $g^2_{YM}\to
\infty$. The covariant derivatives for bi-fundamental matters are  
\begin{eqnarray}
&& D_\mu Z^A = \partial_\mu Z^A + iA_\mu^{(1)}Z^A - i Z^A A_\mu^{(2)},
\\
&& D_\mu W^A = \partial_\mu W^A + iA_\mu^{(2)}W^A - i W^A A_\mu^{(1)}.
\end{eqnarray}
Our normalization is
${\rm tr} [T^a T^b] = \frac12 \delta_{ab}$ for the  $U(N)$ generators 
$T^a$.
The moduli space is 
$({\bf C}^4/{\bf Z}_k)^N/S_N$.
When the CS level $k$ is 1 or 2, 
the supersymmetries are expected to enhance to full ${\cal N}=8$. 
We will not treat fermions in this paper, for simplicity.

\subsection{Orbifold to $S^1$ compactification}
\label{section2-1}

We turn on a vev
for one of the scalar fields, say, $Z^1$:
\begin{eqnarray}
 Z^1 = v {\bf 1}_{N\times N}
\label{vevz}
\end{eqnarray}
where $v$ is real and positive, and ${\bf 1}_{N\times N}$ 
is $N\times N$ unit
matrix. The vevs of the other scalar fields are set to zero.
Basically, $v$ measures how far it is from the orbifold fixed point to
the coincident $N$ M2-branes.\footnote{
This can be seen from the moduli space metric
of the moduli space ${\bf C}^4/{\bf Z}_k$, which is flat if measured by
this $v$.
} 
Note that $Z$ has dimension $1/2$, 
so the distance is given by 
$v l_{\rm P}^{3/2}$ where $l_{\rm P}$ is the Planck length in 
11d M-theory. The ABJM action describes a low energy limit 
$l_{\rm P}\rightarrow 0$ of the $N$ M2-branes with the transverse 
target space ${\bf C}^4/{\bf Z}_k$.

It was discussed in \cite{Mukhi:2008ux}
that taking a large value of the vev $v$ is equivalent
to obtaining a system of multiple D2-branes.\footnote{
Precisely speaking, this limit is to consider $F\ll k^2/v^4$, as
will be explained in (\ref{kvzero}).
}  
This was a first example of non-Abelian duality in 3 dimensions,
as one can trade the non-Abelian degrees of freedom of the adjoint
scalar field ${\rm Im}Z^1$ with its dual non-Abelian gauge field
$A_\mu$.
The elimination of the scalar field promotes the CS gauge
field to a dynamical YM gauge field. 

The discussion of \cite{Mukhi:2008ux} was somewhat mysterious, as there
seems to be no M-theory circle to make a reduction from M-theory to the
Type IIA string theory. This problem was clarified in 
\cite{Distler:2008mk} by taking the limit (\ref{limvk}).
In this limit, the orbifold angle gets smaller 
as the location of the M2-branes is translated far away from the
orbifold fixed point, while the distance from the M2-branes to their
orbifold copy is fixed to be $2\pi v l_{\rm P}^{3/2}/k$. 
Since in the limit the orbifold fixed point is very far away from the
M2-brane location, this is effectively the same as the standard $S^1$ 
compactification. 

This is a clever way to create (by hand) a compactification circle  
by a scaling limit of an orbifold which breaks translational
invariance. 
In the limit of shrinking the circle radius, 
the M2-brane system is expected
to reduce to the system of $N$ D2-branes. This was explicitly shown in 
\cite{Distler:2008mk}: the BLG model in this limit 
reduces to a 3d SYM, the effective action of the D2-branes. 

In \cite{Distler:2008mk}, the BLG model does not describe $N$ M2-branes, 
so we shall use the ABJM model. 
Next we demonstrate, in the limit
(\ref{limvk}), how the ABJM model
with $U(N)\times U(N)$ gauge group reduces to the $U(N)$ SYM,
as an exercise for later convenience.

\subsection{ABJM to 3d YM}
\label{section2-2}

The expectation value (\ref{vevz}) inserted to the scalar kinetic terms in
(\ref{abjm}) produces mass terms for the gauge fields. The scalar fields
are in the bi-fundamental representation, so we choose the following
redefined gauge fields: 
\begin{eqnarray}
 A_\mu^{(\pm)} \equiv \frac12 \left(A_\mu^{(1)}\pm A_\mu^{(2)}\right).
\label{pmdef}
\end{eqnarray}
The mass term arising from the scalar kinetic term is 
\begin{eqnarray}
 S_{\rm mass} =
- \int\! d^3x \; {\rm tr} \left[
\{A_\mu^{(-)}, v\}^2
\right]
= - \int\! d^3x \; 4v^2 {\rm tr} \left[
(A_\mu^{(-)})^2
\right].
\label{gmass}
\end{eqnarray}
In terms of  (\ref{pmdef}), the CS terms in the ABJM
action (\ref{abjm}) are written as 
\begin{eqnarray}
&&
\epsilon^{\mu\nu\lambda}{\rm tr}
\left[
A^{(1)}_\mu \partial_\nu A^{(1)}_\lambda
-A^{(2)}_\mu \partial_\nu A^{(2)}_\lambda
\right]
=  \epsilon^{\mu\nu\lambda}{\rm tr}
\left[
A^{(+)}_\mu \partial_\nu A^{(-)}_\lambda
\right],
\\
&&
\epsilon^{\mu\nu\lambda}{\rm tr}
\left[
A^{(1)}_\mu A^{(1)}_\nu A^{(1)}_\lambda
-A^{(2)}_\mu A^{(2)}_\nu A^{(2)}_\lambda
\right]
= 2 \epsilon^{\mu\nu\lambda}{\rm tr}
\left[
3A^{(+)}_\mu A^{(+)}_\nu A^{(-)}_\lambda
+A^{(-)}_\mu A^{(-)}_\nu A^{(-)}_\lambda
\right],
\nonumber
\end{eqnarray}
up to a total derivative. Then, the CS terms are
\begin{eqnarray}
 S_{\rm CS} = \int d^3x\frac{k}{2\pi}
\epsilon^{\mu\nu\lambda}{\rm tr}
\left[
A^{(-)}_\mu F^{(+)}_{\nu\lambda} + \frac{2i}{3} 
A^{(-)}_\mu A^{(-)}_\nu A^{(-)}_\lambda
\right],
\label{gpm}
\end{eqnarray}
with the field strength
\begin{eqnarray}
 F^{(+)}_{\nu\lambda} \equiv 
\partial_\nu A^{(+)}_\lambda
-\partial_\lambda A^{(+)}_\nu
+ i[A^{(+)}_\nu, A^{(+)}_\lambda].
\end{eqnarray}

{}From $S_{\rm mass} + S_{\rm CS}$ ((\ref{gpm})+(\ref{gmass})), 
it is obvious that $A_\mu^{(-)}$ is an auxiliary field and can be
integrated out. 
We treat the cubic term in (\ref{gpm}) as a perturbation, as it turns
out to be decoupled in the limit (\ref{limvk}). The equation of motion
for $A_\mu^{(-)}$ is (if the cubic term is neglected)
\begin{eqnarray}
 A_\mu^{(-)} = \frac{k}{16\pi v^2}\epsilon_{\mu\nu\lambda}
F^{(+)\nu\lambda}.
\label{eomaf}
\end{eqnarray}
Substituting this back to the action, we obtain
\begin{eqnarray}
 S = -\int d^3x \frac{k^2}{32\pi^2v^2}{\rm tr}\left[
(F_{\mu\nu}^{(+)})^2\right] + \frac{k^4}{v^6}{\cal O}((F^{(+)})^3).
\label{kvzero}
\end{eqnarray}
We have used 
$\eta^{\mu\nu}\epsilon_{\mu\rho\lambda}\epsilon_{\nu\sigma\tau}
= -(\eta_{\rho\sigma}\eta_{\lambda\tau}-
\eta_{\rho\tau}\eta_{\sigma\lambda})$.
The $F^3$ term in this action is from the cubic interaction
$(A^{(-)})^3$ in (\ref{gpm}), whose coefficient goes to zero 
in the limit (\ref{limvk}). 
So we obtain a 3d YM with a finite 
gauge coupling
\begin{eqnarray}
 \lim_{k,v\rightarrow\infty}
\frac{k^2}{32\pi^2v^2} = \frac{1}{2g_{\rm YM}^2}.
\end{eqnarray}
The important and basic mechanism here is that the CS gauge field is
upgraded to a dynamical YM field through Higgsing one scalar 
field.\footnote{To maintain the total degrees of freedom, one of
the scalar field should go away from the system. Interestingly, one can
find that the kinetic term for the imaginary part of $Z^1$ disappears
with the vev of the real part.} 
We use this mechanism throughout the paper. 

In the above, we substituted the result of the classical
equation of motion 
(\ref{eomaf}) into the action classically. However, this can be
justified fully at the quantum level, because the field which is
eliminated is just and auxiliary field. To be concrete, one can show
that integrating out the field $A_\mu^{(-)}$ in the path integral 
approach is equivalent to just substituting the result of the classical  
equation of motion back to the action. 

\section{Generalized ABJM to 4d YM}
\label{section3}
\setcounter{footnote}{1}

\subsection{Generalized ABJM} 
\label{section3-1}

In order to have D3-branes, we need to compactify M-theory on a shrinking 
torus transverse 
to M2-branes. Instead of 
the circle compactification (\ref{limvk}), 
here we need another circle to make the torus. To gain
another circle, a different orbifold with large order is 
necessary. 
We make use of the generalized ABJM model
which was studied in 
\cite{Benna:2008zy,Terashima:2008ba,Imamura:2008dt} 
for our purpose.
The standard orbifolding action of Douglas and Moore 
\cite{Douglas:1996sw}
has been applied to the ABJM model 
to obtain the generalized action
(ver.~2 of \cite{Terashima:2008ba}). Alternatively,
a Type IIB brane realization leading to 
same theory in the IR limit is given in \cite{Imamura:2008dt}. 
The generalized ABJM model 
is characterized by a longer quiver diagram 
(Fig.~\ref{fig}). It was shown in \cite{Terashima:2008ba,Imamura:2008dt} that 
this theory has a more general 
moduli space ${\bf C}^4/\Gamma$, as expected.

In this section, we show that, in a similar 
limit of the orbifold and expectation values of the scalar fields,
the generalized  
ABJM model is equivalent classically to a 4d ${\cal N}=4$ SYM 
theory. 

The bosonic part of the generalized ABJM 
action is
\cite{Benna:2008zy,Terashima:2008ba,Imamura:2008dt}\footnote{The 
lagrangian written here is the one described in \cite{Benna:2008zy}. 
We can think of this as case II in 
\cite{Terashima:2008ba}, or the theory of
\cite{Imamura:2008dt} with $n_A=n_B$, in their notations respectively.
The corresponding Type IIB brane configuration, which  was studied 
in \cite{Imamura:2008dt}, has $n$ NS5-branes and $n$ 
$(k,1)$ 5-branes which are placed pairwise, adjacent to one another,
on an $S^1$ which $N$ D3-branes are wrapping. 
Under the RG flow, 3d CS-matter quiver
gauge theory (\ref{abjm2}) appears at the IR fixed point.}: 
\begin{eqnarray}
S = \int \! d^3x \; 
\left[
\frac{k}{4\pi}\epsilon^{\mu\nu\lambda}
\sum_{l=1}^{n}
{\rm tr}
\left(
A^{(2l-1)}_\mu \partial_\nu A^{(2l-1)}_\lambda + \frac{2i}{3}
A^{(2l-1)}_\mu
A^{(2l-1)}_\nu A^{(2l-1)}_\lambda 
\right.
\right.
\nonumber \\
\left.
-
A^{(2l)}_\mu \partial_\nu A^{(2l)}_\lambda - \frac{2i}{3}A^{(2l)}_\mu
A^{(2l)}_\nu A^{(2l)}_\lambda 
\right)
\nonumber \\
\left.
-{\rm tr} \sum_{s=1}^{2n}\left(
(D_\mu Z^{(s)})^\dagger D^\mu Z^{(s)}
+(D_\mu W^{(s)})^\dagger D^\mu W^{(s)}
\right) - V(Z,W)
\right]. 
\label{abjm2}
\end{eqnarray}
The definition of the covariant derivative is 
\begin{eqnarray}
&& D_\mu Z^{(2l-1)} = \partial_\mu Z^{(2l-1)}
+ i A_\mu^{(2l-1)}Z^{(2l-1)}
- i Z^{(2l-1)} A_\mu^{(2l)}, \\
&& D_\mu Z^{(2l)} = \partial_\mu Z^{(2l)}
+ i A_\mu^{(2l)}Z^{(2l)}
- i Z^{(2l)} A_\mu^{(2l+1)}, \\
&& D_\mu W^{(2l-1)} = \partial_\mu W^{(2l-1)}
+ i A_\mu^{(2l)}W^{(2l-1)}
- i W^{(2l-1)} A_\mu^{(2l-1)}, \\
&& D_\mu W^{(2l)} = \partial_\mu W^{(2l)}
+ i A_\mu^{(2l+1)}W^{(2l)}
- i W^{(2l)} A_\mu^{(2l)}.
\end{eqnarray}
When $n=1$, this reduces to the original ABJM action. 
The quiver diagram is a simple one shown in Fig.~\ref{fig}. 
It is a standard quiver diagram except for the fact that the sign of the CS level 
is opposite for two adjacent nodes. 

\FIGURE{
\includegraphics[width=10cm]{quiver.eps}
\put(-200,20){$Z^{(2l-1)}$}
\put(-150,20){$Z^{(2l)}$}
\put(-100,20){$Z^{(2l+1)}$}
\put(-200,-10){$W^{(2l-1)}$}
\put(-150,-10){$W^{(2l)}$}
\put(-103,-10){$W^{(2l+1)}$}
\put(-170,-25){$A^{(2l)}$}
\put(-220,-25){$A^{(2l-1)}$}
\put(-75,-25){$A^{(2l+2)}$}
\put(-123,-25){$A^{(2l+1)}$}
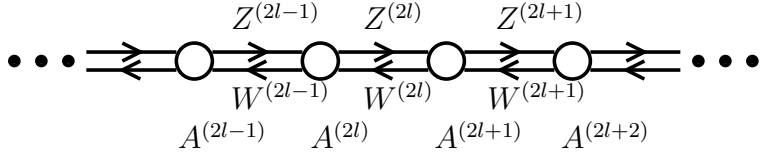
\caption{Quiver diagram of the generalized ABJM model. The quiver 
forms a
 circle with $2n$ nodes.}
\label{fig}
}

The moduli space of this generalized ABJM model is 
${\bf C}^4/({\bf Z}_n \times {\bf Z}_{nk})$ 
\cite{Terashima:2008ba,Imamura:2008dt},
for $N=1$ (a single M2-brane). For general $N$, the moduli space is
$N$ copies of it, $({\bf C}^4/({\bf Z}_n \times {\bf Z}_{nk}))^N/S_N$. 
Due to the parameterization given there, 
the following point in the moduli space, 
\begin{eqnarray}
Z^{(2l-1)} = v 1_{N\times N}, 
\quad  Z^{(2l)}  = \tilde{v} 1_{N\times N}, \quad W^{(2l-1)}
=W^{(2l)}=0
\quad (l=1,\cdots, n),
\label{vev}
\end{eqnarray}
is expected to give a torus compactification, once the limit 
(\ref{limitnn}) is taken.
The values $v/n$ and $\tilde{v}/n$ should be associated with radii
of the transverse $T^2$. We will see this in Sec.~\ref{section4}. 
The torus shrinks to a point in the limit (\ref{limitnn}) 
(while the complex structure is kept), such that
Type IIB string theory dual to M-theory has decompactified 10
dimensions. So we can expect that the limit (\ref{limitnn}) 
will bring the generalized ABJM model to 
a SYM on a decompactified 4 dimensions.

\subsection{Generalized ABJM to 4d YM}
\label{section3-2}

In this subsection, we demonstrate how the 4d YM action
is obtainable from the generalized ABJM model, in the limit 
(\ref{limitnn}), via two steps.\footnote{ 
We focus only on gauge kinetic terms. 
Scalar field parts should be shown in a straightforward manner, so we
will not elaborate on it.} 
\begin{itemize}
 \item 
Consider linear combinations of the
gauge fields labeled by $(+)$ and $(-)$, and then integrate out the
auxiliary field $A^{(-)}$ to obtain the YM kinetic term.
At this stage, the theory is 3-dimensional. 
This step is quite similar to the one considered in 
Sec.\ref{section2-2} for the original ABJM model. 
\item
Accumulate $n$ ($\rightarrow\infty$) YM
fields to form a 4d theory via the familiar field-theoretical
       realization of T-duality formulated by Taylor
       \cite{Taylor:1996ik} and deconstruction of extra dimensions
       \cite{ArkaniHamed:2001ca}.
\end{itemize}
The first step basically corresponds to considering the M-theory circle
to obtain the D2-brane action (though the number of the D2-branes is $n
\rightarrow \infty$ in our case). The second step is for T-dualizing the
D2-branes in the covering space of $S^1$.

\subsubsection{The first step : CS $\rightarrow$ 3d YM}
\label{section3-2-1}

To perform the first step described above, we introduce the following
definition of the linear combination of the gauge fields,
\begin{eqnarray}
 A_\mu^{(\pm)(2l-1)}\equiv \frac12 \left(
A_\mu^{(2l-1)}\pm A_\mu^{(2l)}
\right).
\label{pmgauge}
\end{eqnarray}
Precisely as in Sec.\ref{section2-2}, 
the CS part in (\ref{abjm2}) 
reads with the definition (\ref{pmgauge}) as 
\begin{eqnarray}
&& S = S_{\rm CS} + S_{\rm mass}
\label{abjm3}
\\
&& S_{\rm CS} = \int \! d^3x \! 
\sum_{l=1}^{n}
\frac{k}{2 \pi}\epsilon^{\mu\nu\lambda}
\sum_{l=1}^{n}
{\rm tr}
\left[
A^{(-)(2l\!-\!1)}_\mu F^{(+)(2l\!-\!1)}_{\nu\lambda} 
\!+\! \frac{2i}{3}
A^{(-)(2l\!-\!1)}_\mu
A^{(-)(2l\!-\!1)}_\nu A^{(-)(2l\!-\!1)}_\lambda 
\right].
\nonumber 
\end{eqnarray}

With the vev (\ref{vev}), we get the mass term from 
(\ref{abjm2}) (scalar fluctuations are neglected)
\begin{eqnarray}
S_{\rm mass} &=&
-\int\! d^3x \sum_{l=1}^{n}{\rm tr}
\left[
v^2 (A_\mu^{(2l-1)}-A_\mu^{(2l)})^2 + 
\tilde{v}^2 
(A_\mu^{(2l)}-A_\mu^{(2l+1)})^2
\right].
\nonumber \\
&=&
-\int\! d^3x\sum_{l=1}^{n}{\rm tr}
\left[
4v^2 (A_\mu^{(-)(2l-1)})^2 
\right.
\nonumber \\
& & 
\left.
\hspace{10mm}+ 
\tilde{v}^2 
\left(
(A_\mu^{(+)(2l-1)}-A_\mu^{(+)(2l+1)})
-(A_\mu^{(-)(2l-1)}+A_\mu^{(-)(2l+1)})
\right)^2
\right].
\label{gaugemass2}
\end{eqnarray}
For our later purpose we define mass matrices as
\begin{eqnarray}
&& S_{\rm mass} = \int \! d^3x 
\sum_{l,l'=1}^n
{\rm tr}
\left[
A_\mu^{(-)(2l-1)} M_{ll'}^{(-)} A^{(-)(2l'-1)\mu}
\right.
\nonumber \\
&&
\hspace{20mm}
\left.
+A_\mu^{(-)(2l-1)} M_{ll'}^{({\rm cross})} A^{(+)(2l'-1)\mu}
+A_\mu^{(+)(2l-1)} M_{ll'}^{(+)} A^{(+)(2l'-1)\mu}
\right].
\label{mass+def}
\end{eqnarray}
For reproducing (\ref{gaugemass2}), we define
\begin{eqnarray}
&&
M^{(-)}\equiv -4(v^2+\tilde{v}^2)
 {\bf 1}_{n\times n} + 2 \tilde{v}^2\Lambda,
\quad
M^{({\rm cross})} \equiv 2\tilde{v}^2(\Omega - \Omega^{-1}),
\quad
M^{(+)} \equiv (-\tilde{v}^2) \Lambda, 
\nonumber 
\\
&&
\Lambda \equiv
2 {\bf 1}_{n\times n} -(\Omega + \Omega^{-1}).
\label{massmat}
\end{eqnarray}
The matrix $\Omega_{ij} \equiv \delta_{i+1,j}$ 
is the standard $n\times n$ shift matrix, with 
the indices in the definition $\delta_{i+1,j}$ should be
understood mod $n$. ${\bf 1}_{n\times n}$
is the unit matrix of the size $n\times n$.

It is clear that $A^{(+)(2l-1)}$ is just an auxiliary
field and can be integrated out. 
The mass term (\ref{gaugemass2}) is a little complicated, so in this
subsection we consider a simplified situation
\begin{eqnarray}
 v \gg \tilde{v}.
\end{eqnarray}
In the next subsection we deal with generic $v$ and 
$\tilde{v}$. For $v \gg \tilde{v}$, we can neglect 
$\tilde{v}^2 (A^{(-)})^2$ and the cross terms 
$\tilde{v}^2 A^{(-)}A^{(+)}$. 
Furthermore, as in Sec.\ref{section2-2}, we can ignore
$(A^{(-)})^3$ 
term because it vanishes when $v\rightarrow\infty$
after $A^{(+)}$ is integrated out. 
The action simplifies to 
\begin{eqnarray}
&& S = \int \! d^3x \; 
\sum_{l=1}^{n}
{\rm tr}
\left[
\frac{k}{2 \pi}\epsilon^{\mu\nu\lambda}
\left(
A^{(-)(2l-1)}_\mu F^{(+)(2l-1)}_{\nu\lambda} 
\right)
\right.
\nonumber \\
&& 
\hspace{30mm}
-
\left.
4v^2 (A_\mu^{(-)(2l-1)})^2 
-\tilde{v}^2 
\left(
A_\mu^{(+)(2l-1)}-A_\mu^{(+)(2l+1)}
\right)^2
\right].
\end{eqnarray}
The equation of motion for the auxiliary field $A^{(-)(2l-1)}$ is
\begin{eqnarray}
 A_\mu^{(-)(2l-1)} = \frac{k}{16\pi v^2}\epsilon_{\mu\nu\lambda} 
F^{(+)(2l-1)\nu\lambda},
\label{eoma-}
\end{eqnarray}
and we substitute this back to the action\footnote{
As described at the end of the previous section, this substitution of
the classical equation of motion of the auxiliary fields can be
justified at the quantum level. In the path-integral formalism,
first one shifts the auxiliary field $A^{(-)}$ by the amount 
(\ref{eoma-}) to absorb the CS terms, and then 
can integrate out this shifted auxiliary field because it is decoupled
from the rest. The resultant action is the same as (\ref{massiveym}).
} to obtain a 3d {\it massive} YM action 
\begin{eqnarray}
&& S = \int \! d^3x \; 
{\rm tr}
\left[
-\frac{k^2}{32 \pi^2 v^2} 
\sum_{l=1}^{n}
\left(F_{\mu\nu}^{(+)(2l-1)}\right)^2
+\sum_{l,l'=1}^{n}A_\mu^{(+)(2l-1)} M_{ll'} A^{(+)(2l'-1)\mu}
\right], 
\label{massiveym}
\end{eqnarray}
with $M=M^{(+)}$.
By virtue of (\ref{eoma-}), the three kinds of terms are neglected
safely due to the order estimation:
\begin{eqnarray}
&&\tilde{v}^2 (A^{(-)})^2 \sim 
\tilde{v}^2 v^{-4}
(F^{(+)})^2 \ll v^{-2}
(F^{(+)})^2, \quad
\tilde{v}^2 A^{(-)}A^{(+)} \sim 
\tilde{v}^2 v^{-2}
A^{(+)}F^{(+)},\quad
\nonumber
\\
&& (A^{(-)})^3 \sim v^{-6} (F^{(+)})^3.
\end{eqnarray}

\subsubsection{The second step : 3d YM $\rightarrow$ 4d YM}
\label{section3-2-2}

The 3d massive YM action (\ref{massiveym}) 
which we obtained is, in fact, the one used
for deconstruction \cite{ArkaniHamed:2001ca}. So,
in the limit $n\rightarrow\infty$, the action
(\ref{massiveym}) becomes a 4d YM action.
In the following we will demonstrate this explicitly, in a
self-contained manner. 
In Sec.~\ref{section3-3} we will treat generic values of $v$ and 
$\tilde{v}$, where the complete action is different from 
(\ref{massiveym}) and so needs explicit formulas of deconstruction
for the analysis.\footnote{
Note that deconstruction is, in the limit $n\rightarrow\infty$, the
same as Taylor's field theoretical T-duality, essentially (see Appendix
\ref{app}). This is 
because the orbifold action creating the quiver can be identified as a
circle compactification action. The Taylor's T-duality mainly
concentrates on the scalar part of the theory while deconstruction
treats mostly the gauge field part instead. In this paper we give the
details for the gauge field part of the action and deconstruction. The scalar part should be straightforwardly incorporated
in the same manner.}

To clarify the physical meaning of the action (\ref{massiveym}),
let us diagonalize its mass term. 
The mass spectrum can be seen in the eigenvalues of the mass 
matrix $M^{(+)}$. 
It is well known that the shift matrix can be diagonalized to 
a clock matrix 
$\tilde{\Omega} \equiv {\rm diag}(q,q^2,\cdots, q^{n-1},1)$, with
$q \equiv \exp[2\pi i /n]$.
So the eigenvalues of $\Lambda$ are
\begin{eqnarray}
 \lambda_l = 2-(q^l+q^{-l})= 2-2\cos
\left(\frac{2\pi l}{n}\right)
\end{eqnarray}
where 
$l=[n/2]-n+1, \cdots, -1,0,1,\cdots, [n/2]$
(the range of $l$ is shifted for later convenience).
More precisely, there exists an orthogonal matrix $O$ with which 
we redefine the gauge fields as 
\begin{eqnarray}
 A_\mu^{(+)(2l-1)} = \sqrt{n}O^l_{\; l'}\hat{A}_\mu^{(+)(l')}.
\label{redefA}
\end{eqnarray}
The inclusion of the front factor $\sqrt{n}$ is for our later
convenience. 
Then the diagonalized mass matrix is 
\begin{eqnarray}
O^{\rm T} M^{(+)} O = {\rm diag} (\lambda_{[n/2]-n}, \cdots, 
\lambda_{-1}, \lambda_0, \lambda_1, \cdots, \lambda_{[n/2]}).
\end{eqnarray}
 
In the limit $n\rightarrow \infty$, this mass formula around
the massless level becomes
\begin{eqnarray}
 \lambda_s = (2 s \pi/n)^2 \quad (-\infty < s < \infty, \;
s\in {\bf Z}).
\label{forml}
\end{eqnarray}
the action (\ref{massiveym}) can readily be rewritten 
as
\begin{eqnarray}
 S 
&=& \int\! d^3x 
\; {\rm tr}\left[-
\frac{nk^2}{32\pi^2v^2}
\hat{\cal L}_{\rm kin} 
-4\tilde{v}^2 n\sum_{s\in {\bf Z}} \left(\frac{s\pi}{n}\right)^2
(\hat{A}_\mu^{(+)(s)})^2
\right],
\label{resultac}
\end{eqnarray}
with
\begin{eqnarray}
&& \hat{\cal L}_{\rm kin}
\equiv 
\sum_s(\partial_\mu \hat{A}_\nu^{(+)(s)}
-\partial_\nu \hat{A}_\mu^{(+)(s)})^2
\nonumber \\
&&
\hspace{10mm}
+ 2i\sqrt{n} \sum_{s,s',s'',s'''} O^s_{\;s'} O^s_{\;s''} O^s_{\;s'''}
(\partial_\mu \hat{A}_\nu^{(+)(s')}
-\partial_\nu \hat{A}_\mu^{(+)(s')})
[\hat{A}^{(+)(s'')\mu},\hat{A}^{(+)(s''')\nu}]
\nonumber \\
&&
\hspace{10mm}
 - n\sum_{s,s',s'',s''',s''''} 
O^s_{\;s'} O^s_{\;s''} O^s_{\;s'''}O^s_{\;s''''}
[\hat{A}_\mu^{(+)(s')},\hat{A}_\nu^{(+)(s'')}]
[\hat{A}^{(+)(s''')\mu},\hat{A}^{(+)(s'''')\nu}].
\quad\quad
\label{lkinhat}
\end{eqnarray}
Let us show the final result (\ref{resultac}) signifies the appearance
of a bunch of D3-branes at low energy, {\it i.e.}~a YM action 
{\it in 4 dimensions} compactified on a circle.  
The gauge kinetic term of the 4d YM action is
\begin{eqnarray}
 c \int d^3x d\tau \; {\rm tr} F_{MN}^2, \quad
F_{MN} \equiv \partial_M A_N
-\partial_N A_M + i[A_M, A_N].
\label{acd3}
\end{eqnarray}
The indices run for 4
dimensional coordinates, $M,N=0,1,2,\tau$. The Kaluza-Klein (KK) reduction on
the $S^1$ parameterized by $\tau$ 
can be achieved by Fourier decomposition
\begin{eqnarray}
 A_\mu (x,\tau)= \sum_{s=-\infty}^{\infty} e^{i s \tau/R} 
B_\mu^{(s)}(x). 
\label{fourier2}
\end{eqnarray}
The radius of the circle is $R$.
For simplicity, we neglect the scalar field $A_\tau$.
Substituting this decomposition back to the YM action (\ref{acd3})
and integrating it over $\tau$,
we obtain
\begin{eqnarray}
2\pi R c \int\! d^3x \; 
{\rm tr}\left[
{\cal L}_{\rm kin}
+2 \sum_{s=-\infty}^\infty
\left(\frac{s}{R}\right)^2 B_\mu^{(s)}B^{(-s)\mu}
\right],
\label{d3red}
\end{eqnarray}
with
\begin{eqnarray}
&& {\cal L}_{\rm kin}
\equiv 
\sum_s
(\partial_\mu B_\nu^{(s)}
-\partial_\nu B_\mu^{(s)})
(\partial_\mu B_\nu^{(-s)}
-\partial_\nu B_\mu^{(-s)})
\nonumber \\
&&
\hspace{10mm}
+ 2i\sum_{s'+s''+s'''=0} 
(\partial_\mu B_\nu^{(s')}
-\partial_\nu B_\mu^{(s')})
[B^{(s'')\mu},B^{(s''')\nu}]
\nonumber \\
&&
\hspace{10mm}
 - \sum_{s'+s''+s'''+s''''=0} 
[B_\mu^{(s')},B_\nu^{(s'')}]
[B^{(s''')\mu},B^{(s'''')\nu}].
\quad\quad
\label{lkinhat2}
\end{eqnarray}

Let us show that our action (\ref{resultac}) is indeed equal to the 
KK reduced YM action (\ref{d3red}), with an
appropriate choice of the overall normalization $c$.
For the computation, we need to use an 
explicit expression for the orthogonal matrix $O$.
The eigenvectors of the matrix $\Lambda$ are
\begin{eqnarray}
 V^{(s)} = (1,q^s,q^{2s}\cdots,q^{(n-1)s})^{\rm T}/\sqrt{n}
\label{eigenvec}
\end{eqnarray}
which are labeled by $s=[n/2]-n,\cdots,-1,0,1,\cdots,[n/2]$.
These vectors are orthogonal to each other, due to $q^n=1$.
$O$ is formed by alignment of ortho-normal vectors. 
But vectors (\ref{eigenvec}) are not real-valued, so
in order to make a real-valued matrix $O$ one need to rearrange the
vectors, 
\begin{eqnarray}
 \left(
V'^{(-s)},  V'^{(s)}
\right)
Q=
 \left(
V^{(-s)}, V^{(s)}
\right), \quad 
Q \equiv \left(
\begin{array}{cc}
i/\sqrt{2} & -i/\sqrt{2}  \\ 1/\sqrt{2}  & 1/\sqrt{2} 
\end{array}
\right).
\label{Q}
\end{eqnarray}
Then, $\{V'\}$ is a set of ortho-normalized real vectors, 
forming the matrix $O$ by their alignment. 

However, the
vectors $V$ are simpler than $V'$, so we choose a new basis for the
gauge fields, rather than (\ref{redefA}),
\begin{eqnarray}
  A_\mu^{(+)(2l-1)} \equiv \sqrt{n}(OQ)^l_{\; l'} B_\mu^{(l')}.
\label{newdefA}
\end{eqnarray}
In comparison to our previous basis (\ref{redefA}), 
this is equivalent to 
$B_\mu^{(0)}=\hat{A}_\mu^{(+)(0)}$
and 
\begin{eqnarray}
 B_\mu^{(l)} = \frac{1}{\sqrt{2}}
(\hat{A}_\mu^{(+)(l)} + i \hat{A}_{\mu}^{(+)(-l)}),\quad
 B_\mu^{(-l)} = \frac{1}{\sqrt{2}}
(\hat{A}_\mu^{(+)(l)} - i \hat{A}_{\mu}^{(+)(-l)})\quad
(l>0).
\label{relaba}
\end{eqnarray}
Then, from (\ref{eigenvec}) and (\ref{Q}), we obtain a simple formula
\begin{eqnarray}
 \sqrt{n} (OQ)^l_{\; l'} = q^{ll'}
\label{explicitq}
\end{eqnarray}
and can use it for evaluating $\hat{\cal L}_{\rm kin}$
(\ref{lkinhat}). Using the equations
\begin{eqnarray}
&&
\sum_{s,s',s'',s'''} (OQ)^s_{\;s'} (OQ)^s_{\;s''} (OQ)^s_{\;s'''} 
= n^{-3/2} \sum_{s',s'',s'''}
\sum_{s=[n/2]-n}^{[n/2]}
q^{s(s'+s''+s''')} 
\nonumber \\ 
&& \quad\qquad
= 
n^{-3/2} 
\left[
\sum_{s'+s''+s'''\neq 0}q^{([n/2]-n)(s'+s''+s''')}
\frac{1-q^{n(s'+s''+s''')}}{1-q^{s'+s''+s'''}} 
+
\sum_{s'+s''+s'''=0} n
\right]
\nonumber \\ 
&& \quad\qquad
= 
n^{-1/2} 
\sum_{s'+s''+s'''=0} 1,
\\
&&
\sum_{s,s',s'',s'''} (OQ)^s_{\;s'} (OQ)^s_{\;s''} 
(OQ)^s_{\;s'''} (OQ)^s_{\;s''''} 
= 
n^{-1} \hspace{-8mm}
\sum_{s'+s''+s'''+s''''=0} 1,
\end{eqnarray}
where $q^n=1$ is taken into account,
we can show
\begin{eqnarray}
\hat{\cal L}_{\rm kin}(\hat{A}) = {\cal L}_{\rm kin}(B),
\label{kineq}
\end{eqnarray}
{\it i.e.}, the kinetic term (\ref{lkinhat}) is equal to the
KK kinetic term (\ref{lkinhat2}).\footnote{
The constraints appearing in the sum, $s'+s''+s'''=0$ and 
$s'+s''+s'''+s''''=0$, can be 
interpreted as a momentum conservation in the
$\tau$ space. The physical reason for this is that basically
the element $q=\exp[2\pi i/n]$ is a generator of the clock matrix, 
meaning a translation to the next orbifold copy in the covering space of
the orbifold. 
On the other hand, in the KK expansion
(\ref{fourier2}), the expansion unit is $\exp[i\tau/R]$ which can be
written as an operator $\exp[iP_\tau \tau]$ which is a translation by
the amount of $\tau$ where $P_\tau$ is a conjugate momentum and thus
generate the translation. So, it is natural to identify this
translation with the orbifold translation
in the covering space, $q^s$. Furthermore, in our limit
$n\rightarrow\infty$, the exponent $s/n$ of this 
$q^s = \exp[2\pi i s/n]$ becomes continuous, which is interpreted as
$\tau$. The sum over $s$ means an integration over $\tau$. 
So, we understand that the basis $B_\mu^{(l)}$ (\ref{newdefA})
for the 3d quiver gauge
theory is the KK basis of the 4d YM theory, 
realizing an explicit deconstruction. 
}

Having checked the equivalent structure of the kinetic term, 
we proceed to determine the coefficient $c$ of the 4d
YM action (\ref{acd3}). With use of (\ref{relaba}) and 
(\ref{kineq}), our action (\ref{lkinhat}) is written 
in terms of the fields $B_\mu^{(s)}$ as
\begin{eqnarray}
 S 
&=& -\frac{nk^2}{32\pi^2v^2}
\int\! d^3x \; 
{\rm tr}\left[
{\cal L}_{\rm kin}
+\frac{128\pi^4 v^2 \tilde{v}^2}{k^2n^2}
\sum_s s^2
\left[B_\mu^{(s)}B^{(-s)\mu}\right]
\right].
\label{ourlag}
\end{eqnarray}
Compared with the KK reduced action (\ref{d3red}), 
the compactification circle radius of the 4d theory can be
identified as
\begin{eqnarray}
 \frac{1}{R} = \frac{8\pi^2 v \tilde{v}}{kn}.
\label{rad4}
\end{eqnarray}
Furthermore, comparing the front coefficients of (\ref{d3red}) and
(\ref{ourlag}) shows
\begin{eqnarray}
 2\pi R c = -\frac{nk^2}{32\pi^2v^2}.
\end{eqnarray}
This fixes the constant $c$, so 
finally we find that our
action (\ref{massiveym}) is equal to
\begin{eqnarray}
 S = -\frac{k\tilde{v}}{8\pi v}
\int d^4x \; {\rm tr}\left[F_{MN}^2\right].
\label{ym4re}
\end{eqnarray}
This is a 4d YM action, with the normalization
completely fixed.

The radius $R$ of the $S^1$ in 4 dimensions (\ref{rad4}) 
diverges in our limit (\ref{limitnn}), so we
recover the full YM action in a non-compact 
4d space. We have shown that the generalized ABJM model in the
limit (\ref{limitnn}) is equivalent to the 4d YM
theory (\ref{ym4re}).
The gauge
coupling of the 4d YM theory is given by 
\begin{eqnarray}
\frac{1}{g_{\rm YM}^2} = \frac{k\tilde{v}}{4\pi v}.
\label{gaugec}
\end{eqnarray}  

\subsection{Full 4d action with $\theta$ term}
\label{section3-3}

The action (\ref{ym4re}) obtained in the previous subsection
is insufficient for our purpose, since MO duality 
uses arbitrary value of the gauge coupling. 
In this subsection we derive the 4d theory
for arbitrary values of $v$ and $\tilde{v}$. Here we quote our result
in advance: 
\begin{eqnarray}
 S = 
\int\! d^4x \; {\rm tr}\left[
-\frac{kv\tilde{v}}{8\pi (v^2+ \tilde{v}^2)}F_{MN}^2
+ \frac{k\tilde{v}^2}{16 \pi (v^2 + \tilde{v}^2)}
\epsilon^{MNPQ}F_{MN}F_{PQ}
\right].
\label{ym4recom}
\end{eqnarray}
Interestingly, there appears a $\theta$ term. 
coefficients are finite in the limit (\ref{limitnn}).
The final 4d action (\ref{ym4recom}) is of course consistent
with the previous one (\ref{ym4re}) in the approximation 
$v\gg \tilde{v}$.

\subsubsection{YM term}
\label{section3-3-1}

Our action before assuming $v\gg \tilde{v}$ is (\ref{abjm3}) with the
mass term defined by (\ref{massmat}). We shall follow the steps 
developed in the previous subsection, while keeping all terms.

The equation of motion for the auxiliary field $A_\mu^{(-)(2l-1)}$ is 
\begin{eqnarray}
 A_\mu^{(-)(2l-1)} = 
-\frac{k}{4\pi}\epsilon_{\mu\nu\lambda} ((M^{(-)})^{-1})^l_{\; l'}
F^{(+)(2l'-1)\nu\lambda}
- \frac12 ((M^{(-)})^{-1}(M^{({\rm cross})})^{\rm T})^l_{\; l'}
A_\mu^{(+)(2l'-1)}.
\nonumber
\\
\label{eomaf2}
\end{eqnarray}
Compared with the previous (\ref{eoma-}) 
(which is for $v\gg \tilde{v}$), we have the
additional second term in the right hand side.
We substitute this back\footnote{
 As noted before, this procedure can be justified at the quantum
level. 
} to the action (\ref{abjm3}), to obtain 
\begin{eqnarray}
&&
 S = \int \! d^3x
\; {\rm tr}
\left[
-\eta^{\mu\mu'}\left(
\frac{k}{4\pi}\epsilon_{\mu\nu\lambda} 
F_{\nu\lambda}^{(+)(2l-1)}
+ \frac12 A_\mu^{(+)(2l'-1)}(M^{({\rm cross})})_{l'}^{\;\; l}
\right) 
((M^{(-)})^{-1})_{ll''}
\right.
\nonumber \\
&&
\hspace{40mm}
\times
\left(
\frac{k}{4\pi}\epsilon_{\mu'\nu'\lambda'} 
F^{(+)(2l''-1)\nu'\lambda'}
+ \frac12 ((M^{({\rm cross})})^{\rm T})^{l''}_{\; l'''}
A_{\mu'}^{(+)(2l'''-1)}
\right) 
\nonumber \\
&&
\left.
\hspace{60mm}
+A_\mu^{(+)(2l-1)} M_{ll'}^{(+)} A^{(+)(2l'-1)\mu}
\right].
\label{abjm4}
\end{eqnarray}
This is different from (\ref{massiveym}) in two aspects; (i) 
There is
a new contribution to the $A^{(+)}$ mass term, coming from the first
term in (\ref{abjm4}). (ii)
The cross term in the first term in (\ref{abjm4})
gives rise to a CS coupling 
${\rm tr}[A^{(+)}F^{(+)}]$. The first fact (i) provides a  
modification of the KK mass for the YM theory (Sec.~\ref{section3-3-1}), 
and the second fact (ii) gives rise to a $\theta$ term in 4 dimensions
(Sec.~\ref{section3-3-2}).

The total mass matrix $M$ (as defined in (\ref{massiveym}))
for $A^{(+)}$ is now
\begin{eqnarray}
 M = M^{(+)}
-\frac14 M^{({\rm cross})} (M^{(-)})^{-1}(M^{({\rm cross})})^{\rm T}.
\end{eqnarray}
Noting that all $M^{({\rm cross})}$, 
$(M^{({\rm cross})})^{\rm T}$ and $M^{(-)}$ are written by $\Omega$ and
$\Omega^{-1}$, we can change the ordering of the multiplication as 
\begin{eqnarray}
M^{({\rm cross})} (M^{(-)})^{-1}(M^{({\rm cross})})^{\rm T}
=(M^{(-)})^{-1}M^{({\rm cross})} (M^{({\rm cross})})^{\rm T}.
\end{eqnarray}
Then, using a formula 
$M^{({\rm cross})} (M^{({\rm cross})})^{\rm T} =
4\tilde{v}^4(4\Lambda-\Lambda^2)$, 
we find that in fact the basis (\ref{redefA}) 
can diagonalize the total mass matrix $M$
also in the present case.

We are interested in nearly massless levels in the large $n$ limit, 
so only the lowest order in $\Lambda$ is necessary. Since 
$M^{(-)} = -4(v^2 + \tilde{v}^2) + {\cal O}(\Lambda)$, we find
\begin{eqnarray}
 M &=& (-\tilde{v}^2)\Lambda - \tilde{v}^4 
\frac{1}{-4(v^2 + \tilde{v}^2)} 4\Lambda + {\cal O}(\Lambda^2)
\;=\; 
\frac{-v^2\tilde{v}^2}{v^2 + \tilde{v}^2}\Lambda 
+ {\cal O}(\Lambda^2).
\label{modifm}
\end{eqnarray}
So, in the large $n$ limit,
the difference from the previous case ($v\gg \tilde{v}$) is merely the
definition of the mass matrix: $M$ of (\ref{modifm}) instead of
$M^{(+)}$. 
Looking at our previous result (\ref{resultac}) for the action, we
arrive at the expression after the diagonalization, 
\begin{eqnarray}
 S 
&=& \int\! d^3x 
\;{\rm tr}\left[
\sum_{s}\frac{-nk^2}{32\pi^2(v^2+\tilde{v}^2)}
\hat{\cal L}_{\rm kin} 
-4
\frac{nv^2\tilde{v}^2}{v^2 + \tilde{v}^2}
 \sum_{s} \left(\frac{s\pi}{n}\right)^2
(\hat{A}_\mu^{(+)(s)})^2
\right].\quad 
\end{eqnarray}
Note that not only the mass term but also the normalization of the
kinetic term is changed due to $M^{(-)}$. 

Then, as before, we obtain the dual radius, which happens to be the same
as (\ref{rad4}). With this, we get the 4d YM action
\begin{eqnarray}
 S = \frac{-k^2}{32\pi^2(v^2+\tilde{v}^2)} \frac{1}{2\pi R}
\int d^4x \; {\rm tr}\left[F_{MN}^2\right]
= \frac{-kv\tilde{v}}{8\pi (v^2 + \tilde{v}^2)}
\int d^4x \; {\rm tr}\left[F_{MN}^2\right].
\label{YMcomp}
\end{eqnarray}
This is the first term of our full result (\ref{ym4recom}).

\subsubsection{$\theta$ term}
\label{section3-3-2}

Next, we compute the CS term coming from the cross terms in
the multiplication in (\ref{abjm4}). It is well-known that a dimensional
reduction of a $\theta$ term in 4d YM theory is a
CS term in 3 dimensions. We shall see that this extends to our case.
The CS term of ours is the cross term in (\ref{abjm4}),
\begin{eqnarray}
 S_{\rm cross}=-\int\! d^3x \; \frac{k}{4\pi}
\epsilon^{\mu\nu\lambda}
{\rm tr}
\left[
A_{\mu}^{(+)(2l-1)} (M^{({\rm cross})}
(M^{(-)})^{-1})_{ll'}F_{\nu\lambda}^{(+)(2l'-1)}
\right].
\label{cscross}
\end{eqnarray}
The matrix $(M^{(-)})^{-1}$ 
can be replaced by $(-4(v^2+\tilde{v}^2))^{-1}$ as
before, for nearly-massless levels. However, the matrix 
$M^{({\rm cross})}$ 
cannot be diagonalized by the orthogonal rotation $O$. 
To evaluate this explicitly, again we use the basis of $B_\mu^{(l)}$
(\ref{newdefA}). We obtain
\begin{eqnarray}
&& S_{\rm cross}=\int\! d^3x \; 
\frac{k}{16\pi(v^2\! +\! \tilde{v}^2)}
\epsilon^{\mu\nu\lambda}
{\rm tr}
\left[
(OQ)^l_{\; l'}
B_{\mu}^{(l')} M^{({\rm cross})}_{ll''} 
\left(
n(OQ)^{l''}_{\; l'''} (\partial_\nu B_\lambda^{(l''')}
-\partial_\lambda B_\nu^{(l''')})
\right.
\right.
\nonumber \\
&&
\hspace{50mm}
\left.\left.
+n\sqrt{n} i(OQ)^{l''}_{\; l'''} (OQ)^{l''}_{\; l''''} 
[B_\nu^{(l''')},B_\lambda^{(l'''')}]
\right)
\right].
\label{BBB}
\end{eqnarray}
Using (\ref{explicitq}) and 
$M^{({\rm cross})}_{ij}
=2\tilde{v}^2(\delta_{i+1,j}-\delta_{i-1,j})$, we obtain
\begin{eqnarray}
&&\frac{1}{2\tilde{v}^2}
n (OQ)^l_{\; l'} M^{({\rm cross})}_{ll''} (OQ)^{l''}_{\; l'''}
= \sum_{l',l''} (q^{ll'}q^{(l+1)l'''}-q^{ll'}q^{(l-1)l'''})
\nonumber \\
&& \quad\qquad
=(q^{l'''}-q^{-l'''}) \sum_{l}q^{l(l'+l''')}
= (q^{l'''}-q^{-l'''}) n \delta_{l'+l''',0}
= -4\pi i l' \delta_{l'+l''',0},
\nonumber \\
&& 
\frac{1}{2\tilde{v}^2} n\sqrt{n} (OQ)^l_{\; l'} 
M^{({\rm cross})}_{ll''} (OQ)^{l''}_{\; l'''}
(OQ)^{l''}_{\; l''''}
= -4\pi i l' \delta_{l'+l'''+l'''',0}.
\label{csd}
\end{eqnarray}
These formulas are used to evaluate (\ref{BBB}) to get
\begin{eqnarray}
&& S_{\rm cross}=-\int\! d^3x \; 
\frac{ik\tilde{v}^2}{2(v^2 + \tilde{v}^2)}
\epsilon^{\mu\nu\lambda}
\left[
\sum_{l'} l'
{\rm tr}
\left(
B_{\mu}^{(l')} 
(\partial_\nu B_\lambda^{(-l')}
-\partial_\lambda B_\nu^{(-l')})
\right)
\right.
\nonumber \\
&&
\hspace{50mm}
\left.
+ \sum_{l'+l'''+l''''=0}
il'{\rm tr}
\left(
B_{\mu}^{(l')} [B_\nu^{(l''')},B_\lambda^{(l'''')}]
\right)
\right].
\label{csb}
\end{eqnarray}
If we use a partial integration and a Jacobi identity,
all of these terms vanish. However, in view of the fact that
we have an infinite sum, those procedures may be invalid, so
we keep these terms. 

On the other hand, the $\theta$ term in the 4d YM action is 
\begin{eqnarray}
 S_{\theta}= c'\int \! d^3x d\tau \; {\rm tr}
\left[F_{MN}F_{PQ}\epsilon^{MNPQ}\right].
\label{thetater}
\end{eqnarray}
The Fourier decomposition (\ref{fourier2}) leads to 
\begin{eqnarray}
 S_\theta &=& -4c'\int \! d^4x \;\epsilon^{\mu\nu\lambda}
{\rm tr}
\left(
\partial_\tau A_\mu F_{\nu\lambda}
\right)
\nonumber \\
&=&
 8\pi  c'\int d^3x \;
\epsilon^{\mu\nu\lambda}
\left[
-i\sum_{l} l\;
{\rm tr}
\left(
B_{\mu}^{(l)} 
(\partial_\nu B_\lambda^{(-l)}
-\partial_\lambda B_\nu^{(-l)})
\right)
\right.
\nonumber \\
&&
\hspace{50mm}
\left.
+ \sum_{l+l'+l''=0}
l\; {\rm tr}
\left(
B_{\mu}^{(l)} [B_\nu^{(l')},B_\lambda^{(l'')}]
\right)
\right].
\end{eqnarray}
We obtained
the same structure as our cross term action (\ref{csb}).
Comparing the coefficients, we conclude that
(\ref{csb}) is equal to a 4d $\theta$ term, 
\begin{eqnarray}
 S_{\rm cross} = 
\frac{k\tilde{v}^2}{16 \pi (v^2 + \tilde{v}^2)}
\int\! d^4x \; {\rm tr}\left[
\epsilon^{MNPQ}F_{MN}F_{PQ}
\right].
\end{eqnarray}
This is the second term of (\ref{ym4recom}).
 
Together with (\ref{YMcomp}), we have shown finally 
that the YM action
with a $\theta$ term, (\ref{ym4recom}), is equivalent to our 
generalized ABJM action (\ref{abjm2}), in the limit (\ref{limitnn}).

\subsection{Summary}

The procedures we use, which were explained so far, can be understood as
an equivalence among path-integrated partition functions as follows.
\begin{eqnarray}
&&\int \!\left[\prod_{l=1}^{2n} {\cal D}A^{(l)}\right]
 e^{iS_{\rm generalized \; ABJM}}
=\int \!\left[\prod_{l=1}^{n} {\cal D}A^{(+)(2l-1)}
\prod_{l=1}^{n} {\cal D}A^{(-)(2l-1)}
\right] e^{iS_{\rm generalized \; ABJM}}
\nonumber \\
&&
=\int \!\left[\prod_{l=1}^{n} {\cal D}A^{(+)(2l-1)}
\right] e^{iS_{\rm 3d \;  massive\; YM}}
=\int \!\left[\prod_{l=1}^{n} {\cal D}\hat{A}^{(+)(l)}
\right] e^{iS_{\rm 3d \;  massive\; YM}}
\nonumber \\
&&
=\int \!{\cal D}A_{(4d)}
\; e^{iS_{\rm 4d \;  YM}}.
\end{eqnarray}
The first equality is just a field redefinition by a linear
combination (\ref{pmgauge}). 
At the second equality, we integrated out the auxiliary
fields $A^{(-)(2l-1)}_\mu$. This was explained with the substitution of
the classical equation of motion (\ref{eomaf2}), but it can be justified
at the quantum level. 
At the third equality, we rotate the basis
of the gauge field labels as in (\ref{redefA}), 
and so it is merely a linear 
field redefinition. At the last equality, we sum up the KK tower and
rewrite the action just in 4d terminology.

As is obvious from this
equality, the generalized ABJM model (which was our starting point) 
and
the 4d YM theory are equivalent to each other at the quantum level.
Of course one can show this equivalence in the presense of field
operators in the path integrals, so equivalence among correlators can be
shown.
Note that the action is considered as a bare action
of the path-integral 
with an appropriate cut-off.\footnote{
At this point there is a subtlety about taking the infinite cut-off limit.
However, in our case the supersymmetry of the 3d action will 
constrain the action and we do not expect any problem for it.}

\section{$SL(2,{\bf Z})$ duality}
\label{section4}
\setcounter{footnote}{1}

We have obtained the 4d YM theory (\ref{ym4recom})
from the generalized ABJM model (\ref{abjm2}). 
The 4d YM action  (\ref{ym4recom}) has a
complexified gauge coupling
\begin{eqnarray}
 \tau = \frac{-k \tilde{v}^2}{v^2 + \tilde{v}^2}
+ i \frac{k v\tilde{v}}{v^2 + \tilde{v}^2}
\label{tau}
\end{eqnarray}
where $\tau$ is of the standard notation, 
\begin{eqnarray}
&& S = \frac{-1}{8\pi}\int \! d^4x \; 
{\rm tr}
\left[
{\rm Im}(\tau)
F_{MN}F^{MN}
\!+\! {\rm Re}(\tau)
\frac12 \epsilon^{MNPQ}F_{MN}F_{PQ}
\right],
\quad
\tau \equiv \frac{\theta}{2\pi} + \frac{4\pi i}{ g_{\rm YM}^2}.
\nonumber 
\end{eqnarray}
In this section, we
first show that in fact from the single theory (\ref{abjm2}) we can
obtain infinite number of 4d YM theories
(\ref{ym4recom}) which differ in values of $\tau$
(Sec.~\ref{section4-1}). 
This explicitly proves equivalence between these 4d
theories. Indeed we will show that all of these theories are related to
each other by $SL(2,{\bf Z})$ transformations 
and the pariy transformation (Sec.~\ref{section4-2}). 
Finally in Sec.~\ref{section4-3}, a consistent 
interpretation in M-theory and superstring theory is given.

\subsection{Infinitely many equivalent 4d theories}
\label{section4-1}

In the previous derivation, we have chosen a linear
combination (\ref{pmgauge}) of the gauge fields, then one of the
combinations become the auxiliary field $A^{(-)}_\mu$ and is integrated
out. Note that we may have other choice of the linear combination. 
In fact, for a
gauge field $A_\mu^{(2l-1)}$ with the CS level $k$, we have
$n$ choices for $A_\mu^{(2l')}$ with $-k$, to form a linear 
combination.

As a typical example, let us choose the following new combination:
\begin{eqnarray}
 A_\mu^{(\pm)(2l-1)} \equiv \frac12 (A_\mu^{(2l-1)}\pm A_\mu^{(2l-2)}). 
\label{newdeflin}
\end{eqnarray}
Here the labels are understood with mode $2n$, 
{\it i.e.}~$A^{(0)}=A^{(2n)}$, $A^{(-1)}=A^{(2n-1)}$. Apparently, with
this new 
basis all the computations in the previous section can be done as well. 
The only difference is the 
exchange of $v$ and $\tilde{v}$. In fact, with the definition
(\ref{newdeflin}), the mass term for the gauge field is 
\begin{eqnarray}
 && S_{\rm scalar}
=-\int\! d^3x \; \sum_{l=1}^{n}{\rm tr}
\left[
4\tilde{v}^2 (A_\mu^{(-)(2l-1)})^2 
\right.
\nonumber \\
&& 
\left.
\hspace{20mm}+ 
v^2
\left(
(A_\mu^{(+)(2l-1)}-A_\mu^{(+)(2l+1)})
-(A_\mu^{(-)(2l-1)}+A_\mu^{(-)(2l+1)})
\right)^2
\right],
\label{gaugemass3}
\end{eqnarray}
while the CS kinetic term in (\ref{abjm3}) is left intact.
The resultant 4d YM 
action has a coupling constant 
\begin{eqnarray}
 \tau' 
= \frac{-k v^2}{v^2 + \tilde{v}^2}
+ i \frac{k v\tilde{v}}{v^2 + \tilde{v}^2},
\label{tau'}
\end{eqnarray}
which is obtained just with $v\leftrightarrow \tilde{v}$ on the original
coupling constant (\ref{tau}).

Note that we did not modify the generalized ABJM action itself: 
what we changed is just
the labeling of the gauge fields. We are dealing with an identical
theory. So the YM with $\tau$ (\ref{tau}) is equivalent to
the YM with $\tau'$ (\ref{tau'}).

We may choose other combinations for the gauge fields. Next, we consider
an example
\begin{eqnarray}
 A_\mu^{(\pm)(2l-1)} \equiv \frac12 (A_\mu^{(2l-1)}\pm A_\mu^{(2l+2)}). 
\label{lin2}
\end{eqnarray}
This combination provides a complicated
mass term for the gauge fields. In terms of the definition of the mass
matrix (\ref{mass+def}), the linear combination (\ref{lin2}) leads to
\begin{eqnarray}
 M^{(-)} &=& -v^2 ({\bf 1}+ \Omega + \Omega^{-1})
-\tilde{v}^2 (2{\bf 1}+ \Omega^2 + \Omega^{-2}),\\
M^{({\rm cross})} &=& 2v^2(\Omega-\Omega^{-1}) +
2\tilde{v}^2(\Omega^2-\Omega^{-2}),\\
 M^{(+)} &=& -v^2 (2{\bf 1}- \Omega - \Omega^{-1})
-\tilde{v}^2 (2{\bf 1}- \Omega^2  \Omega^{-2}).
\end{eqnarray}
With these mass matrices, the computations presented in
Sec.~\ref{section3-3} can be done quite similarly, and we arrive at a
4d YM theory with
\begin{eqnarray}
 \tau' = \frac{-k (v^2 + 2\tilde{v}^2)}{v^2 + \tilde{v}^2}
+ i \frac{k v\tilde{v}}{v^2 + \tilde{v}^2}.
\label{tau''}
\end{eqnarray}
This theory is, again, equivalent to the YM theory with
(\ref{tau}) and also to the one with (\ref{tau'}).

In this manner, we can continue choosing different combinations. 
A generalization of the combination (\ref{lin2}) is
\begin{eqnarray}
 A_\mu^{(\pm)(2l-1)} \equiv \frac12 (A_\mu^{(2l-1)}\pm A_\mu^{(2l+2m)})
\end{eqnarray}
for arbitrary positive integer $m$ ($m<n$), and for each choice
we arrive at a different value of $\tau$.
In the end, 
we obtain infinite number of various gauge coupling constants for
the 4d YM theory, all of which are equivalent.
Next, let us see how these coupling constants are related to each other.

\subsection{$SL(2,{\bf Z})$ relation}
\label{section4-2}

MO duality group for $U(N)$ 
${\cal N}=4$ YM
theory is $SL(2, {\bf Z})$, and we here show that the relation between
the original $\tau$ and the infinite variety of $\tau'$ is indeed 
given by this transformation. The 
$SL(2, {\bf Z})$ transformation is
\begin{eqnarray}
 \tau' = \frac{a \tau + b}{c\tau + d}, \quad ad-bc=1, \quad
a,b,c,d\in {\bf Z}.
\end{eqnarray}

First, we consider possible relation between (\ref{tau}) and
(\ref{tau''}). 
We substitute (\ref{tau}) and (\ref{tau'}) into the above
and seek for a solution for 
the integer set $(a,b,c,d)$ satisfying $ad-bc=1$. 
In terms of the standard notation for the generators of the 
$SL(2,{\bf Z})$ group : the shift operation  
$T(\tau) = \tau +1$ and the
inversion $S(\tau)=-\tau^{-1}$, we find
\begin{eqnarray}
 \tau' = \tau-k = T^{-k}(\tau).
\label{t}
\end{eqnarray}
So, the choice (\ref{lin2}) 
of the linear combination for the gauge fields realizes
the $T$-transformation of the $SL(2,{\bf Z})$ group.
This is quite interesting and encouraging: The different pairing of the
CS gauge fields results in an $SL(2,{\bf Z})$-transformed complexified gauge
coupling! \footnote{Although the $T$-transformation is a generic
symmetry of the gauge theory, we stress that 
the M-theory torus and its $SL(2,{\bf Z})$ group action
is behind our realization of the $T$-transformation.}

Any realization of the $T$-transformation in the $SL(2,{\bf Z})$ group 
is nontrivial. It is often stated in literature that $T$-transformation
is trivial because one can easily see the invariance of the partition
function under the transformation: the $\theta$ term couples to the
instanton number which is quntized, so the $T$ shift of the $\theta$
angle changes the value of the action by $2\pi$ which leaves any 
path integration invariant. However, to the best of our knowledge,
nobody has realized this shift by a transformation of the fields.
Our method concretely realizes this transformation of the fields, as 
a change of the pairings of the CS gauge fields in the KK-reduced 3
dimensions.

Then how about the $S$-transformation which is more interesting in the
MO duality? For this, let us look at a relation between (\ref{tau})
and (\ref{tau'}).
In fact, there is a solution for $(a,b,c,d)$ for 
$k=1$ and $k=2$,
\begin{eqnarray}
 (a,b,c,d)= (-1,-1,2,1) \quad (\mbox{for}\; k=1) \\
 (a,b,c,d)= (-1,-2,1,1) \quad (\mbox{for}\; k=2) 
\end{eqnarray}
which is equivalent to
\begin{eqnarray}
&& \tau' = S(T^2(S(T(\tau)))) 
\quad (\mbox{for}\; k=1) \label{s1}\\
&& \tau' = T^{-1}(S(T(\tau))) 
\quad (\mbox{for}\; k=2) \label{s2}
\end{eqnarray} 
Note that these include the inversion $S$.

With these facts presented, can we claim that MO duality
is proved? The answer is NO. 
Note that our coupling
constant (\ref{tau}) is not generic. It is parameterized by one real
parameter $v/\tilde{v}$, so the 4d YM theory we
obtained probes only a small portion of the 
fundamental domain of $SL(2,{\bf Z})$.
We find that this is fatal in respect of the MO duality.
It turns out that a combination of (\ref{s1}) (or (\ref{s2})) with the
other one (\ref{t}) is equivalent to a parity transformation in 4
dimensions. In fact, the combination leads to\footnote{More precisely,
this $\tau'$ can be obtained by considering a linear combination basis 
$A_\mu^{(\pm)(2l-1)} \equiv \frac12 (A_\mu^{(2l)}\pm A_\mu^{(2l+1)})$.
With this choice, previous computations can be
performed only with the exchange of $k\rightarrow -k$. In this paper we
have assumed that $k$ is positive. If we allow for arbitrary sign for
$k$, then our formula for $\tau$ (\ref{tau}) is 
$\tau = (-k \tilde{v}^2 + i |k v\tilde{v}|)/(v^2 + \tilde{v}^2)$. So the
change of the sign of $k$ flips the sign of the real part of $\tau$.}
\begin{eqnarray}
 \tau' = 
 \frac{k \tilde{v}^2}{v^2 + \tilde{v}^2}
+ i \frac{k v\tilde{v}}{v^2 + \tilde{v}^2}
\end{eqnarray}
which is only different in the sign of the $\theta$, compared to the
original $\tau$ (\ref{tau}). This is a parity
transformation in 4 dimensions. 

Note that this $\tau'$ can also be represented as 
a combination of $S$- and $T$-transformations. In other words, 
the $S$-transformation (\ref{s1}) or (\ref{s2}) which we realized 
by a totally field-theoretical argument for a proof of MO duality
can also be obtained by a parity transformation in 4 dimensions. 

Our one-parameter family of $\tau$ 
lies on parts of the boundary of the fundamental domain of 
$SL(2,{\bf Z})$ (we choose a conventional definition of the
fundamental domain). The parts of the boundaries are identified by 
some of the $SL(2,{\bf Z})$ transformations, and in our case eventually
this transformation can also be understood as a parity transformation.
This is a peculiarity of our coupling constant (\ref{tau}). 
For generic values of the gauge coupling constant, the parity
transformation would not be 
equivalent to any $SL(2,{\bf Z})$ transformation.\footnote{
The conventional choice of the fundamental domain is defined by 
a region in $\tau$ complex plane given by $|\tau|\geq 1$, 
$-1/2 \leq {\rm Re} \tau \leq 1/2$ 
and ${\rm Im} \tau>0$. If we consider the
parity as well as $SL(2,{\bf Z})$, the whole moduli space of the
4d YM theory is a half of the fundamental domain defined
above; one needs to further restrict it to the region 
${\rm Re}\tau\geq 0$. In this moduli space, our coupling constant $\tau$ 
lies on fixed lines of the ``parity + $SL(2,{\bf Z})$'' duality group.
}
%

We present another fact. In (\ref{s1}) and (\ref{s2}) we have chosen
$k=1,2$. However, with other choice of the value of $k$, we cannot find 
$SL(2,{\bf Z})$ transformation $\tau\to\tau'$. On the other hand, if we 
allow the parity transformation, $\tau$ and $\tau'$ can be related
for any $k$. Even though we can choose arbitrary $k$ for giving infinite
variety of the values of the coupling constant $\tau$ via the CS
pairings, this fact suggests that we had better understand this $\tau'$
as a parity, rather than $S$-transformation, generically.\footnote{
Although this choice of $k=1,2$ is special in the sense that 
it leads to the full supersymmetry ${\cal N}=8$ in 3 dimensions for the
original ABJM model (see \cite{Aharony:2008ug}).}

\subsection{M-theory interpretation}
\label{section4-3}

In this paper, so far, we have used only traditional techniques 
of field theories, and haven't used any technologies 
of string theory and M-theory. 
But the reason why we got a particular value of $\tau$ 
(\ref{tau}) will be clear once
string theory interpretation is used, as we will see.
As described in the introduction, $\tau$ can be identified
with the torus modulus $\tau$ for the compactification of 11-dimensional
M-theory. We made this torus by turning on the scalar vevs $v$ and
$\tilde{v}$ and taking the limit (\ref{limitnn}). The modulus of the
torus is associated with the scalar vevs through the orbifolding action 
which can be seen in the moduli space of multiple M2-branes.  

\FIGURE{
\includegraphics[width=5cm]{torus.eps}
\put(-110,130){${\rm Im} \; z^2$}
\put(-10,70){${\rm Im} \; z^1$}
\put(-35,10){$\vec{v}_1$}
\put(-145,90){$\vec{v}_2$}
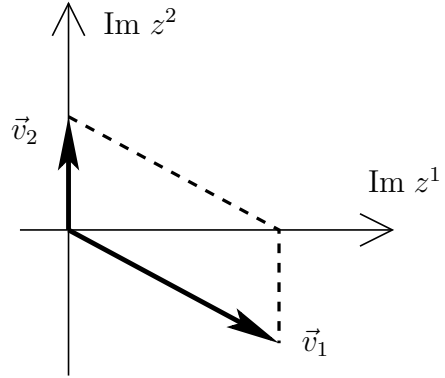
\caption{Transverse torus made by the limiting orbifold.}
\label{fig2}
}
In our case, the orbifold charge acting on the four complex scalar
fields is (see case II of \cite{Terashima:2008ba} or
\cite{Imamura:2008dt}) 
\begin{eqnarray}
 \left(
\frac{1}{kn}, -\frac{1}{kn}, -\frac{1}{kn}, \frac{1}{kn}
\right), \;
 \left(
0,0,\frac{1}{n}, -\frac{1}{n}
\right).
\label{oc}
\end{eqnarray}
This means that the identification is 
\begin{eqnarray}
&& (z_1,w_1,z_2,w_2) 
\nonumber 
\\
&& \sim
(e^{2\pi i/kn}z_1,
e^{-2\pi i/kn}w_1,e^{-2\pi i/kn}z_2,e^{2\pi i/kn}w_2)
\nonumber \\
&&\sim  
(z_1,
w_1,e^{2\pi i/n}z_2,e^{-2\pi i/n}w_2).
\end{eqnarray}
We turned on a vev for the scalar field corresponding to the first and
the third entries. The torus cycles are defined by the circles made by
the limit of the orbifold. The vector that defines the cycles 
of the torus
can be read from the vev vector 
$(z_1,w_1,z_2,w_2)=(v,0,\tilde{v},0)$ and the orbifold
charge (\ref{oc}). For the second charge vector in (\ref{oc}), it is
obvious that the torus cycle direction is (see (\ref{circle}) in the 
appendix for an explicit relation between the standard circle
compactification and a scaling limit of an orbifold)
\begin{eqnarray}
\vec{v}_2\equiv (0,0,2\pi i\tilde{v}/n,0).
\end{eqnarray}
In the same manner, 
from the first vector in (\ref{oc}), another cycle 
vector of the torus is 
\begin{eqnarray}
 \vec{v}_1\equiv (2\pi iv/kn,0,-2\pi i\tilde{v}/kn,0).
\end{eqnarray}
Therefore, defining a complex coordinate made out of the imaginary parts
of the first first ${\bf C}$ and the third ${\bf C}$, we can write 
the vectors $\vec{v}_{1,2}$ giving the cycles of the torus in terms of a
complex coordinate (spanned by imaginary parts of the first and the third
${\bf C}$),
\begin{eqnarray}
 v_1 = 2\pi \left(
\frac{v}{kn} - i \frac{\tilde{v}}{kn}
\right), \quad
 v_2 = 2\pi i \frac{\tilde{v}}{n}.
\end{eqnarray}
See Fig.~\ref{fig2}.
The size of the torus shrinks to zero in the limit (\ref{limitnn}),
while the complex structure of the torus made of these two vectors is
finite, 
\begin{eqnarray}
 \tau = v_2/v_1 = 
 \frac{-k \tilde{v}^2}{v^2 + \tilde{v}^2}
+ i \frac{k v\tilde{v}}{v^2 + \tilde{v}^2}.
\end{eqnarray}
So, in the limit (\ref{limitnn}), M-theory is 
compactified on a shrinking torus transverse to the M2-branes,
with the above $\tau$.

Also, via duality chains, 
M2-branes transverse to 
this torus will ultimately become $N$ D3-branes with 
the background axio-dilaton $\tau$. Therefore, our previous result, 
(\ref{ym4recom}), which has the same $\tau$ (\ref{tau}), 
is consistent with this M-theory interpretation. In other words, 
we find that our
resultant action (\ref{ym4recom}) is consistent with the moduli space
analyzed by \cite{Terashima:2008ba} and \cite{Imamura:2008dt}.

\section{Conclusion and discussion}
\label{section5}


{}From the 3d CS-matter theory we constructed  
the 4d ${\cal N}=4$ $U(N)$ supersymmetric Yang-Mills theory.
This provides explicit $T$-transformations of the $SL(2,Z)$ 
duality for the 4d theory. 

This utilizes two field theory techniques.
One is deconstruction
\cite{ArkaniHamed:2001ca} (or equivalently Taylor's field-theoretical
T-duality \cite{Taylor:1996ik}), which relates a 3d YM and a 4d YM (see
Sec.~\ref{section3-2-2}). The other is new, under which a 3d
superconformal CS-matter theory is Higgsed to a 3d YM
\cite{Mukhi:2008ux,Distler:2008mk} (see Sec.~\ref{section2-2} and
Sec.~\ref{section3-2-1}). Equipped with the two, we are able to
transform the 3d CS-matter theory into the 
4d YM (whose action is obtained in
(\ref{ym4recom})) at the lagrangian level. Roughly speaking, the T-duality
involves a scalar vev $\tilde{v}$, while another scalar vev $v$ triggers
the new duality {\it a la} Mukhi {\it et al.}. 

We showed that a ``reparameterization invariance'', which is nothing but
the relabeling of gauge fields, in the CS-matter theory corresponds to 
the $T$-transformation of the resultant 4d YM. One
reparameterization which amounts to exchanging $v$ and $\tilde{v}$
is indeed an $S$-transformation of the $SL(2,Z)$ MO duality in 4d YM. 
However, in
our restricted fundamental domain, the $S$-transformation here can also
be achieved by a 4d parity and $T$-transformations, so it is not a
strong-weak duality. 

At first glance, our procedures are classical, but integrating out the
auxiliary fields can be justified at the quantum level, so our
equivalence among 4d YM theories with various values of the 
coupling constant is a quantum equivalence.

We believe that, since our method indeed realizes 
a part of the $SL(2,{\bf Z})$ duality manifest
from the M-theory viewpoint, it could be generalized further
including $S$-duality,
possibly by investigating membrane actions in M-theory further.
One may feel that
the $T$-transformations, which we reaized in this work,  are trivial, 
as $T$-invariance can be easily seen
in the path-integral formalism. However, there are two reasons why we
think our results for the $T$-transformations nontrivial. First, we
acheived the shift of the $\theta$ angle, not by hand, but by explicit
redefinition/integration of {\it fields}. 
Second, when the spacetime has a boundary, 
instanton numbers in 4d is not quantized, thus the $T$-transformation,
which changes the action, is quite nontrivial. 
Our procedure can work even for spacetimes with boundaries.

Unlike electric-magnetic (EM) 
duality in abelian case, which relies on introducing a
lagrange multiplier for the Bianchi identity (see \cite{BI} for 
abelian Born-Infeld actions), in our case, the proof involves the novel
Higgs mechanism. This is because in order to promote the 3d theory to 4d 
the
infinite KK tower of massive gauge modes is necessary. Since
ABJM model has an explicit stringy setup regardless of the 
gauge group rank,
it is interesting to see how the abelian EM duality using the lagrange
multiplier can
be consistently understood from the viewpoint of our derivation.

In addition, the torus we made is somewhat artificial due
to our specific choice of moduli points such that
${\bf C}^4/({\bf Z}_A \times {\bf Z}_B)$ reduces to ${\bf C}^2/({\bf
Z}_A \times {\bf Z}_B)$. This moduli space is similar to a
$\beta$-deformed ${\bf C}^2$ without B-field and dilaton. As discussed
in \cite{LM}, supported by the very B-field, D3-branes puff up into  
toroidal D5-branes wrapping a fuzzy two torus, known as Myers effect. 
Since our torus contains the M-circle, a codimension two object is
absent.  

Let us end this section with some comments. One is about the
non-locality of the duality. This involves operations like 9-11 flip in
M-theory. As claimed by Susskind \cite{Susskind:1996uh}, in the context
of Matrix theory, the origin of MO duality can be traced to  the
interplay between circles in strongly-coupled 11 dimensions. It would be
interesting to find a possible relation to that. 


In our derivation so far, 
we have not dealt with scalars and fermions. Fermionic sector 
is in particular important to see that the resultant 4d action has
${\cal N}=4$ supersymmetries. 
In Appendix \ref{app2}, we study the fermionic sector and show that
indeed we obtain ${\cal N}=4$ SYM. The important fact is that
supersymmetries are enhanced from the original 8 supercharges in the
generalized ABJM model to 16 supercharges of the 4d ${\cal N}=4$
SYM. This is a consequence of the scaling limit.

So far our derivation is for a one-parameter family within the
fundamental domain of $\tau$. The possibility to find moduli
spaces which exhibit other quiver diagrams may
shed new light on rendering a full $\tau$ for exploring MO duality. This
remains as an important future work.

\acknowledgments 

K.~H.~would like to thank M.~Nitta and H.~Suzuki for valuable comments. 
T.~S.~T. thanks K.~Ohta and M.~Yamazaki for helpful discussions.
S.~T.~thanks F.~Yagi for useful discussions. 
K.~H.~and S.~T.~are partly supported by
the Japan Ministry of Education, Culture, Sports, Science and
Technology. We would like to thank 
the Yukawa Institute for Theoretical Physics at Kyoto University, at
which we discussed this topic during the workshop YITP-W-08-04 on 
``Development of Quantum Field Theory and String Theory''.


\appendix

\section{Taylor's T-duality and orbifold}
\label{app}

The Taylor's field theory T-duality is for a circle compactification,
while ours makes use of a scaling limit of an orbifold. 
In the limit (\ref{limitnn}), we expect that the circle compactification
emerges. We shall see in this appendix that in fact this emergence can
be seen in the orbifolding action.

First, note that 
the 3d 
YM action (\ref{massiveym}) can be thought of as 
a standard quiver YM theory with a vev of all the bi-fundamental
scalar fields. The mass term in (\ref{massiveym}) can be written as  
\begin{eqnarray}
 S_{\rm mass} = -\int \! d^3x \; {\rm tr}
[A^{(+)}_\mu, \Omega \tilde{v}]^2, \quad
A^{(+)}_\mu \equiv {\rm diag}(A_\mu^{(+)(1)}, A_\mu^{(+)(3)},
A_\mu^{(+)(5)},\cdots).
\label{newmass}
\end{eqnarray}
The part $\Omega \tilde{v}$ can be thought of as a vev of a certain
complex scalar field in adjoint representation, which we call $\Phi$, 
that is, $\langle \Phi \rangle = \Omega \tilde{v}$. 
This scalar field of the size $nN\times nN$, 
after the following orbifold projection
\begin{eqnarray}
 \tilde{\Omega}  \Phi \tilde{\Omega}^\dagger = e^{2\pi i/n} \Phi
\label{orT}
\end{eqnarray}
with the clock matrix $\tilde{\Omega}$, 
has components allowed only for nonzero components of $\Omega$.
This results in bi-fundamental matters in the quiver
YM theory \cite{Douglas:1996sw}.
We turned on a vev $\tilde{v}$ for all the nonzero components of $\Phi$,
that is the interpretation of the mass term (\ref{newmass}).
This is the standard orbifolding for YM theory.
Note that this orbifolding can be thought of as 
the orbifolding for the CS gauge fields of the ABJM model
of \cite{Terashima:2008ba} (see also \cite{Fuji:2008yj}). 
So the emergence
of the orbifold structure in (\ref{massiveym}) is quite natural.

Let us see that this interpretation of the theory (\ref{massiveym})
indeed shows the equivalence to the circle compactification.
We consider a field expanded around its expectation value: 
\begin{eqnarray}
\Phi = \Omega 
\tilde{v} + {\rm Re}(\delta \Phi) +i \; {\rm Im}(\delta \Phi).
\end{eqnarray}
Then, we take a limit $\tilde{v},n\rightarrow\infty$ while $\tilde{v}/n$
fixed. The orbifold action (\ref{orT}) reduces to
\begin{eqnarray}
 \tilde{\Omega}
\left[{\rm Re} (\delta \Phi)\right] 
\tilde{\Omega}^\dagger = {\rm Re} \delta \Phi, \quad 
\tilde{\Omega} \left[{\rm Im} (\delta \Phi)\right] 
\tilde{\Omega}^\dagger = {\rm Im} \delta \Phi + 2\pi \frac{\tilde{v}}{n}.
\label{circle}
\end{eqnarray}
This is precisely the discrete 
action of a circle compactification. Note that the standard discrete
action uses the shift matrix $\Omega$ instead of the clock matrix
$\tilde{\Omega}$, but this difference is merely a convention of
the basis of the matrices.
The discrete action 
on a YM theory (with adjoint scalar fields) was studied by
Taylor \cite{Taylor:1996ik} to show the T-duality concretely in terms
of field theories. The
YM theory divided by the action (\ref{circle}) is shown to be
equivalent to a YM theory in a spacetime with one dimension
higher, compactified on an $S^1$ circle. Therefore, in our case, we
conclude that our action (\ref{massiveym}) is equal to the 
4d YM action compactified on an $S^1$. 

\section{Fermionic sector and ${\cal N}=4$ SUSY in 4d}
\label{app2}

In this appendix, we show that the 4-dimensional Yang-Mills action which
we derived indeed has the expected maximal ${\cal N}=4$ supersymmetries
in 4 dimensions. 

The generalized ABJM action
\cite{Benna:2008zy,Terashima:2008ba,Imamura:2008dt} 
in Sec.~\ref{section3} has 8 supercharges 
(${\cal N}=4$ supersymmetries in 3 dimensions). 
The vacuum expectation values (\ref{vev}) do not break these
supersymmetries. In the 4-dimensional terminology, these 8 supercharges
correspond to 
${\cal N}=2$ supersymmetries in 4 dimensions. 
Now, we note the following fact: in 4
dimensions, ${\cal N}=2$ supersymmetric gauge theory with 4 massless
adjoint fermions is in fact ${\cal N}=4$ supersymmetric Yang-Mills theory.
Therefore, in order to show that our 4-dimensional Yang-Mills action has
${\cal N}=4$ supersymmetries, we only need to show that, after the
deconstruction, we have 4 massless adjoint fermions in 4 dimensions.

In the following, we shall show that this is indeed the case. 
Let us consider the fermion sector of the ABJM model,
\begin{eqnarray}
&& S = 
\int\! d^3x
\left[
L^{\rm ferm}_{\rm kin} - V^{\rm ferm}_{D}
- V^{\rm ferm}_{F}
\right],
\\
&& L^{\rm ferm}_{\rm kin} \equiv 
{\rm Tr}\left[
i\zeta^\dagger \gamma^\mu D_\mu \zeta
+i\omega^\dagger \gamma^\mu D_\mu \omega
\right],
\\
&& 
V^{\rm ferm}_{D}
\equiv
\frac{2\pi i}{k}
{\rm Tr}
\left[
\left(\zeta^\dagger_A \zeta^A - \omega_A \omega^{\dagger A}\right)
\left(Z^\dagger_B Z^B-W_B W^{\dagger B}\right)
\right.
\nonumber
\\
&& 
\hspace{50mm}
\left.
-\left(\zeta^A\zeta^\dagger_A-\omega^{\dagger A}\omega_A\right)
\left(Z^B Z^\dagger_B-W^{\dagger B}W_B\right)
\right]
\nonumber
\\
&& 
\hspace{10mm}
+\frac{4\pi i}{k}
{\rm Tr}
\left[
\left(Z^\dagger_A \zeta^A - \omega_A W^{\dagger A}\right)
\left(\zeta^\dagger_B Z^B-W_B \omega^{\dagger B}\right)
\right.
\nonumber
\\
&& 
\hspace{50mm}
\left.
-\left(\zeta^A Z^\dagger_A-W^{\dagger A}\omega_A\right)
\left(Z^B \zeta^\dagger_B-\omega^{\dagger B}W_B\right)
\right],
\\
&&
V^{\rm ferm}_{F}
\equiv
\frac{2\pi}{k}
\epsilon_{AC}\epsilon^{BD}
{\rm Tr}
\left[
2\zeta^A W_B Z^C \omega_D + 
2\zeta^A \omega_B Z^C W_D
+ Z^A \omega_B Z^C W_D + \zeta^A W_B \zeta^C W_D
\right]
\nonumber
\\
&& 
\hspace{5mm}
+\frac{2\pi}{k}\epsilon_{AC}\epsilon^{BD}
{\rm Tr}
\left[
2\zeta_A^\dagger W^{\dagger B} Z^\dagger_C \omega^{\dagger D} 
+ 2\zeta^\dagger_A \omega^{\dagger B}Z_C^\dagger W^{\dagger D}
+ Z_A^\dagger \omega^{\dagger D}Z_C^\dagger \omega^{\dagger D}
+ \zeta^{\dagger}_A W^{\dagger B}\zeta^\dagger_C W^{\dagger D}
\right].
\nonumber
\end{eqnarray}
Here $A,B=1,2$ are indices for doublets in $SU(2)$ R-symmetry.
For the generalized ABJM model
\cite{Benna:2008zy,Terashima:2008ba,Imamura:2008dt} which we used in
Sec.~\ref{section3}, we just need to follow the orbifolding 
procedures of Douglas and Moore \cite{Douglas:1996sw}: First generalize
the matrix size from $N\times N$ to $nN \times nN$, and then restrict
the matrix elements so that they satisfy the orbifold
constraint. Concretely speaking, we substitute the following expression
to the above lagrangian:
\begin{eqnarray}
 Z^1 = v \Omega_{n\times n} \otimes 1_{N\times N},\quad
Z^2 = \tilde{v} 1_{n\times n} \otimes 1_{N\times N},\quad
W^1=W^2=0.
\end{eqnarray}
This is the same as (\ref{vev}). As for the fermions, we label them as
\begin{eqnarray}
&&
 \zeta^1 = 
\left(
\begin{array}{ccccc}
0 & \zeta^{(3)} & & & \\
0 & 0 & \zeta^{(5)} & & \\
& & \ldots & & \\
0 & 0 & & 0 & \zeta^{(2n-1)}\\
\zeta^{(1)} & 0 &  & & 0
\end{array}
\right),
\quad
 \omega^1 = 
\left(
\begin{array}{ccccc}
0 & 0 & & & \omega^{(1)}\\
\omega^{(3)} & 0 &  & & \\
0&\omega^{(5)} & 0 & & \\
 &  & \ldots & & \\
0 & 0 & & \omega^{(2n-1)} & 0
\end{array}
\right), 
\nonumber 
\\[3mm]
&&
\zeta^2 = {\rm diag}
(\zeta^{(2)}, \zeta^{(4)}, \cdots,\zeta^{(2n)}),
\quad
\omega^2 = {\rm diag}
(\omega^{(2)}, \omega^{(4)}, \cdots,\omega^{(2n)}).
\end{eqnarray}
Each $\zeta^{(t)}$ and $\omega^{(t)}$ $(t=1,2,\cdots,2n)$ are
$N\times N$ matrices. Substituting these matrices to the potentials, we
obtain, for the $\zeta$ sector,
\begin{eqnarray}
 V^{\rm ferm}_D + V^{\rm ferm}_F
= \frac{4\pi i}{k}
v \tilde{v}
\sum_{t,t'} \zeta^{(t)}
\left(\Omega_{2n\times 2n} - \Omega_{2n\times 2n}^{-1}
\right)_{tt'}
\zeta^{(t')*}
+ {\rm c.c.}
\label{massf}
\end{eqnarray}
The size of this shift matrix $\Omega$ is $2n\times 2n$
(on the other hand the shift matrix used in Sec.~\ref{section3}
has the size $n\times n$). $\omega$ sector has precisely the same form,
and is decoupled from the $\zeta$ sector.

We diagonalize the mass term (\ref{massf}). The diagonalization
formula obtained by
replacing $n$ in (\ref{csd}) by $2n$ is
\begin{eqnarray}
\left(\Omega_{2n\times 2n} - \Omega_{2n\times 2n}^{-1}
\right)_{tt'}
\rightarrow
\left(q^{t'/2}-q^{-t'/2}\right)
\delta_{t+t',0}
\label{oo}
\end{eqnarray}
with $q\equiv \exp[2\pi i/n]$. In the large $n$ limit, this simplifies
to 
\begin{eqnarray}
\left(q^{t'/2}-q^{-t'/2}\right)
\delta_{t+t',0}
\rightarrow 
\frac{-2\pi i}{n}t \; \delta_{t+t',0}
+\frac{-2\pi i}{n}(n-t) \; \delta_{t+t',0}
\end{eqnarray}
as we look at modes close to the zero mode. 
Note that in the present case there is the second term,
the almost massless modes around $t\sim n$.\footnote{
In the evaluation of (\ref{csd}), this second term can be discarded
because the Yang-Mills term does not give small mass for these
second sequence. But in the present case, there is no other 
term which generate masses, so we need to pick up all the almost-zero
eigenvalues in the matrix (\ref{oo}). This second term is 
a ``doubler'' since the kinetic 
function is $\left(q^{t'/2}-q^{-t'/2}\right)$ which behaves as a sin
function and has two zeros, as in the standard lattice fermions.}
In the large $n$ limit, the sector $t\sim 0$ decouples from 
the other sector $t\sim n$, so, as a consequence, we obtain two
towers of massive states.
These towers corrrespond to $\zeta^1$ and $\zeta^2$ 
because the projection onto $\zeta^1$, {\it i.e.}
diag$(1,0,1,0, \cdots, 1,0)$, 
commutes with $\Omega-\Omega^{-1}$.
Therefore, in the diagonal basis, we obtain two sets of mass terms
\begin{eqnarray}
 \frac{8\pi^2 v \tilde{v}}{kn}
\sum_t
\left(\zeta^{(t)} t \zeta^{(-t)*} 
+\zeta^{(t-n)} (t-n) \zeta^{(-t+n)*} 
\right)
+ {\rm c.c.}
\end{eqnarray}
This is nothing but a KK mass tower of two 4-dimensional massless
fermions
compactified on a circle with the radius
\begin{eqnarray}
 R = \frac{kn}{8\pi^2 v \tilde{v}}.
\end{eqnarray}
This radius is in exact agreement with the radius obtained in the
analysis of the gauge sector, (\ref{rad4}). Together with the $\omega$
sector which produces two 4-dimensional massless
fermions in precisely the same
manner, we obtain four massless fermions in 4 dimensions.

All of these fermions are in the adjoint representation of the gauge
group in 4 dimensions. This can be seen as follows. 
In order to find the representation of the fermion, it is enough 
to see how the fermions are transformed under the global part of the
gauge transformation in 4 dimensions. Among the KK gauge fields
$B_\mu^{(s)}$ in 3 dimensions, the massless one $B_\mu^{(0)}$ is
relevant to the global part of the gauge transformation.
One can see from (\ref{newdefA}) that this massless mode 
is made of a linear combination of $A_\mu^{(+)(2l-1)}$ with equal
weight. In other words, the first column of the matrix $O$ is
proportional to a vector $(1,1,1,\cdots)$. This means that, the global
transformation corresponds to a simultaneous rotation of all $U(N)$'s
by an equal angle. That is, the global transformation of the
4-dimensional Yang-Mills theory is the global part
of the {\it overall} $U(N)$ of the original $U(N)^{2n}$ gauge group
in 3 dimensions.
Under this overall rotation in the generalized ABJM model, 
all fermions are transformed as the adjoint
representation. Therefore, our 4-dimensional fermions are in the adjoint representation.

\newcommand{\J}[4]{{\it #1} {\bf #2} (#3) #4}
\newcommand{\andJ}[3]{{\bf #1} (#2) #3}
\newcommand{\AP}{Ann.\ Phys.\ (N.Y.)}
\newcommand{\MPL}{Mod.\ Phys.\ Lett.}
\newcommand{\NP}{Nucl.\ Phys.}
\newcommand{\PL}{Phys.\ Lett.}
\newcommand{\PR}{ Phys.\ Rev.}
\newcommand{\PRL}{Phys.\ Rev.\ Lett.}
\newcommand{\PTP}{Prog.\ Theor.\ Phys.}
\newcommand{\hep}[1]{{\tt hep-th/{#1}}}

\end{document}